\documentclass[twocolumn]{emulateapj}

\newcommand{\msun}{{\it M}$_{\odot}$}

\newcommand{\etal}{{\it et al.}}
\newcommand{\ie}{{\it i.e.}}
\newcommand{\eg}{{\it e.g.}}
\newcommand{\be}{\begin{equation}}
\newcommand{\ee}{\end{equation}}

\submitted{The Astrophysical Journal 2010  (in press)}

\shorttitle{Fornax Cluster GC Sizes}
\shortauthors{Masters \etal}

\begin{document}

\title{The ACS Fornax Cluster Survey VII: \\ Half-Light Radii of
  Globular Clusters in Early-Type Galaxies} 

\author{Karen L. Masters\altaffilmark{1,2}, 
  Andr\'{e}s Jord\'{a}n\altaffilmark{3,2}, 
  Patrick C\^{o}t\'{e}\altaffilmark{4}, 
  Laura Ferrarese\altaffilmark{4}, 
  John P. Blakeslee\altaffilmark{4},
  Leopoldo Infante \altaffilmark{3}, 
  Eric W. Peng\altaffilmark{5,6},
  Simona Mei\altaffilmark{7}, 
  Michael J. West\altaffilmark{8}}

\altaffiltext{1}{Institute for Cosmology and Gravitation, University
  of Portsmouth, Dennis Sciama Building, Burnaby Road, Portsmouth, PO1
  3FX, UK}

\altaffiltext{2}{Harvard-Smithsonian Center for Astrophysics, 60
  Garden Street, Cambridge, MA 02138, USA}

\altaffiltext{3}{Departamento de Astronomía y Astrofísica, Pontificia
Universidad Cat\'{o}lica de Chile, 7820436 Macul, Santiago, Chile}

\altaffiltext{4}{Herzberg Institute of Astrophysics, Victoria, BC V9E 2E7, Canada}

\altaffiltext{5}{Department of Astronomy, Peking University, Beijing 100871, China}

\altaffiltext{6}{Kavli Institute for Astronomy and Astrophysics, Peking University, Beijing 100871, China}

\altaffiltext{7}{GEPI, Observatoire de Paris, Section de Meudon, 5 Place Jules Janssen, 92195 Meudon Cedex, France}

\altaffiltext{8}{ESO, Alonso de C\^{u}rdova 3107, Vitacura, Santiago, Chile}

\email{karen.masters@port.ac.uk}

\begin{abstract}
We measure the half-light radii of globular clusters (GCs) in 43
galaxies from the ACS Fornax Cluster Survey (ACSFCS). We use these data
to extend previous work in which the environmental dependencies of the
half-light radii of GCs in early type galaxies in the ACS Virgo
Cluster Survey (ACSVCS) were studied, and a corrected mean half-light
radius (corrected for the observed environmental trends) was suggested
as a reliable distance indicator. This work both increases the sample
size for the study of the environmental dependencies, and adds
leverage to the study of the corrected half-light radius as a possible
distance indicator (since Fornax lies at a larger distance than the
Virgo cluster). We study the environmental dependencies of the size of GCs using both a Principal Component Analysis (PCA) as
well as 2D scaling relations.  We largely confirm the
environmental dependencies shown in \citet{J05}, but find evidence
that there is a residual correlation in the mean half-light radius of
GC systems with galaxy magnitude, and subtle differences in the other
correlations --- so there may not be a universal correction for the
half-light radii of lower luminosity galaxy GC systems. The main
factor determining the size of a GC in an early type galaxy is the GC
color. Red GCs have $\langle r_h \rangle = 2.8\pm0.3 $~pc, while blue
GCs have $\langle r_h \rangle = 3.4\pm0.3$~pc. We show that for bright
early-type galaxies ($M_B < -19$ mag), the uncorrected mean half-light
radius of the GC system is by itself an excellent distance indicator
(with error $\sim 11$\%), having the potential to reach cosmologically
interesting distances in the era of high angular resolution adaptive
optics on large optical telescopes.

\end{abstract}


\section{Introduction}

 The launch of the Hubble Space Telescope (HST) revolutionized areas
 of astronomy which push the limits of high resolution imaging. One
 area which benefited has been the study of the half-light radii,
 $r_h$, of globular clusters (GCs). A typical GC has a half-light
 radius of a few parsecs. Ground-based imaging (before the era of
 adaptive optics) could resolve such objects only in
 galaxies in the Local Group (LG). This limited the statistics and
 confined the study to the GC systems in dwarf and late-type
 galaxies (\eg,~ the sample in \citealt{F00}). HST opened the study to GC systems in galaxies out to $\sim
 30$ Mpc, encompassing large numbers of all types of galaxies and
 including both the Virgo and Fornax clusters (at $\sim 16.5\pm1.1$
 Mpc from \citealt{M07} and $20.0\pm 1.4$ Mpc from \citealt{B09},
 respectively).

Interest in the half-light radius of GCs results both from the
constraints they provide on the formation and evolution of GCs and
also their possible use as a distance indicator, which dates back to
an initial suggestion by \citet{SS27}. The half-light radii of GCs,
rather remarkably, are almost independent of GC mass
\citep{M00,J05,Barmby07,McLaugh08,Harris2009,H09b}, at least to $\sim
10^6 M_\odot$ \citep{H05,M09}. In simulations they have been shown to
be fairly constant as the GCs evolve \citep{ST72,LS78,M90,AH98}, and
may in fact trace the characteristic sizes of the proto-GC cloud
\citep{ML92,H09b}. In the Milky Way, it has long been known that $r_h$
increases systematically with galactocentric distance, although HST
studies \citep{KW01,J05,Harris2009} have shown that in early types,
$r_h$ is much closer to being constant with galactocentric radius.

In this paper we extend the work of \citet[hereafter J05]{J05} which
used ACS data on GCs in Virgo cluster galaxies (from the ACS Virgo
Cluster Survey; ACSVCS, \citealt{2004ApJS..153..223C}) to study the
environmental impacts on the $r_h$ of GCs and calibrated a corrected mean
half-light radius (i.e., corrected for the observed environmental trends) as
a distance indicator. We add to this study similar data from the ACS
Fornax Cluster Survey (ACSFCS; \citealt{J07a}). This adds 43 galaxies
to the sample, and also extends the lever arm for calibration of the
corrected mean half-light radius as a distance indicator since the
Fornax cluster is at a larger distance than the Virgo cluster. 

Other recent works have studied the half-light radius of GCs,
extending both the total number and range of morphologies and
environments studied. GC sizes in M31 and NGC~5128 were studied with
emphasis on the fundamental plane of GCs by \citet{Barmby07} and
\citet{McLaugh08}, respectively. \citet{Barmby07} show (using data for
M31, NGC~5128, the MW, the Magellanic Clouds and the Fornax dwarf
spheroidal) that old GCs appear to have near-universal structural
properties. Measurements of GC sizes in late type galaxies beyond the
LG include the Sombrero galaxy \citep{S06, H09b}, NGC~891
\citep{H09a} and NGC~5190 \citep{Forbes10}. Extremes in host galaxy
luminosity are encompassed by the work on dwarf galaxies by
\citet{Georgiev09} and the study of six giant ellipticals of
\citet{Harris2009}. \citet{DG07} studied the SB0 galaxy NGC~1533 in
the Dorado group. At the limit of what can be currently done from the
ground is the work \citet{Gomez07}, who used IMACS at the Magellan
telescopes to study GC sizes in NGC~5128. As we will comment below,
these works extend many of the results on GC sizes we have obtained
using the ACSVCS and ACSFCS to different host galaxy morphologies and
environments. All in all, the structural properties of GCs seem to
share many near-universal properties accross galaxy morphology and
luminosity, but some differences seem to exist  as well.

The paper is organized as follows. In Section 2 we describe the
data. In Section 3 we present the distributions of $r_h$ in Fornax
early type galaxies. In Section 4 we describe the possible
environmental dependencies which we explore with a principle
components analysis (PCA) in Appendix A and traditional 2D trends in
Section 5. In Section 6 we show the final corrections which are
discussed in Section 7, both as tracers of the GC formation and
evolution and for implications on the use of a corrected mean
half-light radius as a distance indicator. Section 8 presents a
summary of our conclusions.

\section{Observations}
As part of the ACSFCS (\citealt{J07a}), a sample of 43 early type
galaxies in the Fornax cluster were observed in the F475W ($\approx$
Sloan $g$) and F850LP ($\approx$ Sloan $z$) filters. 
The ACSFCS sample was constructed from the Fornax Cluster
     Catalog (FCC; \citealt{F89}) as described in \citet{J07a}
     using the FCC galaxy morphologies.
Half-light radii are measured for all GC candidates using the
procedure explained in detail in J05 which fits a PSF-convolved King
model to the GCs in both bands. The final result quoted here is the
average of the measurement in both bands. The measurement of $r_h$ is
obviously challenging in the cases for which the GCs are only marginally
resolved. J05 show in their Appendix~A that their 
code recovers $r_h$ with no bias to $r_h \sim 0.3$ pixels (which is
0.015$\arcsec$ or $\sim 1.2$ pc at the distance of Virgo; $\sim 1.4$
pc at the distance of Fornax) under typical observing conditions,
assuming the PSF model is correct and that GCs are correctly modeled
by a King profile. As discussed in J05, the systematics undertainties
due to the modeling of the PSF are of the order of 0.05 WFC pixels (2.5 milli-arcseconds), or
$\sim 0.25$ pc at the distance of Fornax. The level of systematic
uncertainties was estimated in J05 by comparing the measurements done
independently in each of the two available bands. This
estimate of the level of systematic uncertainty agrees well with
independent estimates of this quantity given by
\citet{Harris2009}, who uses different datasets and code to measure
$r_h$ with ACS/WFC, and with those derived from a comparison of $r_h$
measurements done using two different ACS/WFC datasets of M87 using the same
code we use here \citep{Peng09}. Thus, the $r_h$ we measure, and in
particular their potential use as distance indicator, are not tied
to our particular observational setup beyond possible systematic
effects at the level of a few milli-arcseconds.\footnote{One caveat is
  that the effects of mass segregation can make the measured $r_h$
  wavelength dependent, with an expected variation of $\sim 5\%$
  between the $V$- and $I$-bands (see, e.g., \citet{Madrid09}).}

We follow J05 in the construction of GC catalogs in each galaxy. The
procedure is described in detail in \citet{J09} and summarized in J05,
so we only list here in brief the cuts made. GCs are selected using
the maximum likelihood estimated probability \citep{J09} of $p_{\rm
  GC} \geq 0.5$, $z$-band magnitude, $z \leq 23.35$ mag, and colors,
$0.6 \leq (g - z) \leq 1.7$ mag. This selection on magnitude is
roughly equivalent to the Virgo cut of $z \leq 22.9$ mag at the
expected turn-over of the GC luminosity function shifted to the
greater distance of Fornax. Contamination and reliability cuts result
in a size range of $r_h=0.75$--$10$ pc for objects which can be reliably
identified as GCs.

The ACSFCS measures distances using the surface brightness fluctuation
(SBF) method for all 43 galaxies in the sample \citep{B09}, therefore
allowing study of the {\it physical} half-light radii of all GCs. We
restrict the sample to galaxies with 5 or more GCs (also removing FCC
202 whose GC system is overwhelmed by its much larger companion, FCC
213). This leaves a final sample of 37 Fornax cluster galaxies. Table
\ref{table1} lists these galaxies showing (1) FCC number, (2) B-band
magnitude from Ferguson (1989), (3) number of globular clusters, (4) average value of $r_h$
calculated using a biweight location estimator, (5) alternative names
for the galaxies.
 
\begin{deluxetable}{lcrcl}
\tablewidth{0pc} 
\tablecaption{Basic Information on Galaxies and GC Systems from ACSFCS Sample}
\tablehead{\colhead{Name} & \colhead{B mag} & \colhead{$N_{\rm GC}$} & \colhead{$\langle r_h \rangle$ (arcsec)} & \colhead{Other Name} 
\label{table1}}
\startdata 
 FCC 21 &  ~9.40 & 232 & 0.040$\pm$0.002 & NGC 1316 \\
FCC 213 & 10.60 & 698 & 0.027$\pm$0.001 & NGC 1399 \\
FCC 219 & 10.90 & 220 & 0.029$\pm$0.001 & NGC 1404 \\
NGC 1340 & 11.27 & 137 & 0.030$\pm$0.002 & NGC 1344 \\
FCC 167 & 11.30 & 266 & 0.032$\pm$0.001 & NGC 1380 \\
FCC 276 & 11.80 & 232 & 0.025$\pm$0.001 & NGC 1427 \\
FCC 147 & 11.90 & 201 & 0.028$\pm$0.001 & NGC 1374 \\
IC 2006 & 12.21 &  89 & 0.034$\pm$0.002 & ESO 359-G7 \\
FCC 184 & 12.30 & 216 & 0.030$\pm$0.001 & NGC 1387 \\
 FCC 83 & 12.30 & 173 & 0.032$\pm$0.001 & NGC 1351 \\
 FCC 63 & 12.70 & 142 & 0.034$\pm$0.002 & NGC 1339 \\
FCC 193 & 12.80 &  23 & 0.046$\pm$0.006 & NGC 1389 \\
FCC 170 & 13.00 &  30 & 0.030$\pm$0.004 & NGC 1381 \\
FCC 153 & 13.00 &  28 & 0.037$\pm$0.004 & IC 1963 \\
FCC 177 & 13.20 &  47 & 0.038$\pm$0.003 & NGC 1380A \\
 FCC 47 & 13.30 & 184 & 0.032$\pm$0.001 & NGC 1336 \\
FCC 310 & 13.50 &  18 & 0.028$\pm$0.004 & NGC 1460 \\
 FCC 43 & 13.50 &  15 & 0.036$\pm$0.006 & IC 1919 \\
FCC 190 & 13.50 & 103 & 0.041$\pm$0.002 & NGC 1380B \\
FCC 148 & 13.60 &  13 & 0.035$\pm$0.007 & NGC 1375 \\
FCC 249 & 13.60 & 102 & 0.036$\pm$0.002 & NGC 1419 \\
FCC 255 & 13.70 &  57 & 0.035$\pm$0.003 & ESO 358-G50 \\
FCC 277 & 13.80 &  17 & 0.034$\pm$0.003 & NGC 1428 \\
 FCC 55 & 13.90 &  12 & 0.034$\pm$0.004 & ESO 358-G06 \\
FCC 152 & 14.10 &   5 & 0.050$\pm$0.021 & ESO 358-G25  \\
FCC 143 & 14.30 &  33 & 0.033$\pm$0.004 & NGC 1373 \\
 FCC 95 & 14.60 &  13 & 0.049$\pm$0.011 & MCG-06-08-025\\
FCC 136 & 14.80 &  13 & 0.045$\pm$0.004 & MCG-06-08-027\\
FCC 182 & 14.90 &  22 & 0.039$\pm$0.005 & MCG-06-09-008 \\
FCC 204 & 14.90 &   7 & 0.050$\pm$0.020 & ESO 358-G43 \\
 FCC 90 & 15.00 &   8 & 0.040$\pm$0.010 & MCG-06-08-024\\
 FCC 26 & 15.00 &  14 & 0.041$\pm$0.008 & ESO 357-G25 \\
FCC 106 & 15.10 &   5 & 0.042$\pm$0.020 & - \\
FCC 324 & 15.30 &   7 & 0.036$\pm$0.007 & ESO 358-G66 \\
FCC 100 & 15.50 &   9 & 0.044$\pm$0.013 & - \\
FCC 203 & 15.50 &  10 & 0.046$\pm$0.007 & ESO 358-G42 \\
FCC 303 & 15.50 &  10 & 0.051$\pm$0.009 & MCG-06-09-028
\enddata
\end{deluxetable}

In some sections, we also consider the sample of GCs in Virgo cluster
galaxies discussed in J05. Because of slight changes in the program
which calculated the GC probabilities, $p_{\rm GC}$,  we find
a slightly different sample than is used in J05; in general, we
find that each galaxy has a few more GCs than were used in J05. We
also add the galaxies VCC21, VCC1833, VCC1440 and VCC1075 which have
very small GC systems and previously were just below the cut-off of
$N_{\rm GC} = 5$ used in J05.

\section{Distribution of $r_h$ in Fornax Cluster Galaxies}

In this paper we will use several different symbols to describe the
half-light radius of GCs. We start this section by providing a
reference table of these symbols. Table~\ref{rhsymbols} lists all the
symbols used along with a brief description of what they represent and
a reference to the section of the paper that defines it (if
appropriate).

\begin{deluxetable*}{ll}
\tablewidth{0pc} 
\tablecaption{Symbols Used  to Describe the Half-Light Radii of Globular Clusters.}
\tablehead{\colhead{Symbol} & \colhead{Description} 
\label{rhsymbols}}
\startdata 
$r_h$ & Measured half-light radius for an individual GC \\
$\langle r_h \rangle$ & Average (usually biweight mean) of all measures of $r_h$ in a given galaxy \\
$r'_h$ & $r_h$ corrected for dependence on local galaxy surface brightness (see Section 6.2, Eqn 2) \\
$\langle r'_h \rangle$ & Average of $r'_h$ \\
$r''_h$ & $r'_h$ corrected for dependence of $\langle r'_h \rangle$ on galaxy color (see Section 6.3, Eqns 6 and 7)\\
$\langle r''_h \rangle$ & Average of $r''_h$ \\
$\hat{r}_h$  & $r''_h$ corrected for dependence on GC color (see Section 6.4)\\
$\langle \hat{r}_h \rangle$ & Average of $\hat{r}_h$ \\
$\hat{r}'_h$ & $\hat{r}_h$ corrected for a residual dependence of $\langle r'_h \rangle$ on galaxy magnitude (see Sections 6.3 and 7).\\
$\langle \hat{r}'_h \rangle$ & Average of $\hat{r}'_h$ 
\enddata
\end{deluxetable*}

\begin{figure*}
\epsscale{1}
\plotone{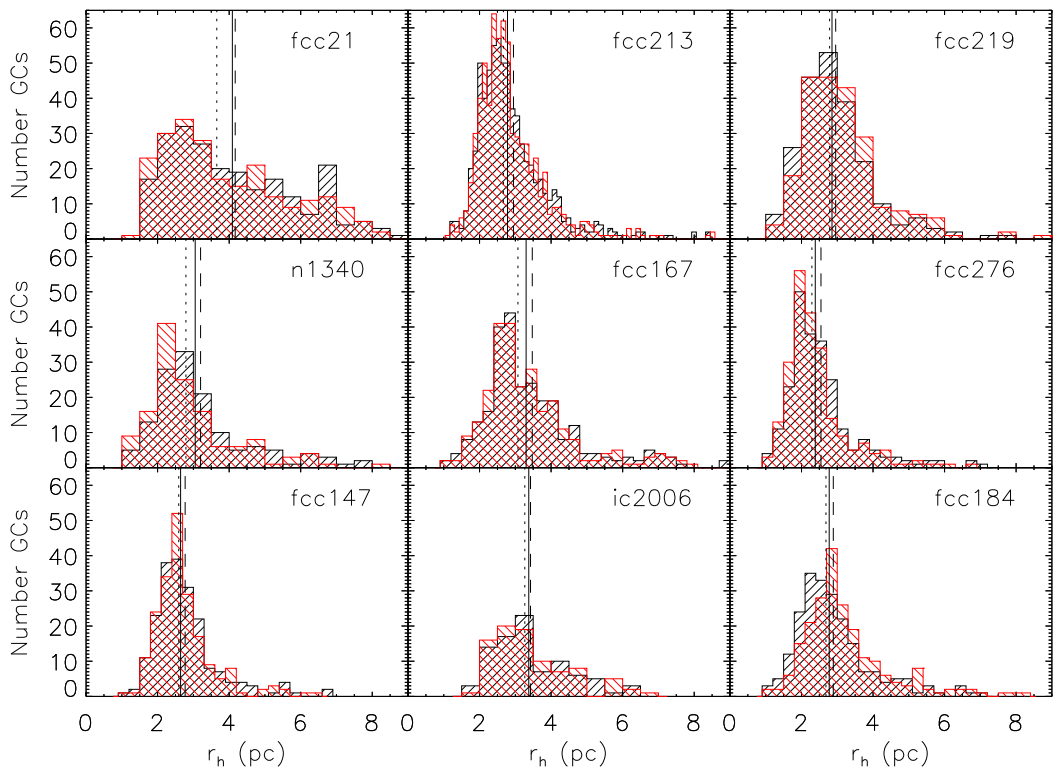}
\caption{The distribution of uncorrected $r_h$, and corrected
$\hat{r}_h$ values (in pc) for the nine most luminous galaxies in
Fornax. Uncorrected values are shown by the forward diagonal hashed
histogram, corrected by the backwards diagonal (red) hashed
histogram. For the uncorrected values, the solid vertical line shows
the biweight location estimation of the mean, the dashed line is the
normal mean, the dotted line is the median value.
\label{rhhists}}
\end{figure*}

The distribution of both raw $r_h$, and corrected ($\hat{r}_h$) values
for all GCs in the system is shown for the nine most luminous Fornax
cluster galaxies in Figure \ref{rhhists}. In those figures, the solid
vertical line shows the biweight location estimation of the mean of
$r_h$, the dashed line is the normal mean and the dotted line is the
median value. The distribution of $r_h$ in all cases is quite far from
normal with a long tail to large values of $r_h$ as was previously
found by J05 for GC systems in the ACSVCS. The median value of $r_h$
is therefore always smaller than the other estimates of the mean. In
addition to galaxies in the Local Group, Virgo, and Fornax, this form
of the GC size distribution has now been observed in galaxies with a
wide range of morphologies and luminosities \citep{Georgiev09,
  Harris2009, H09b}.

FCC 21 (or Fornax A) shows a rather peculiar distribution of GC
half-light radii, with a much larger than normal number of extended
GCs. This skews the average $r_h$ for FCC~21 to a much larger value than
expected. Note that FCC 21 is the most luminous early-type galaxy in the Fornax
cluster and shows unmistakable evidence for a recent merger \citep{S80,G01} which
may account for this odd distribution of GC sizes.\footnote{Recently,
  \citet{daCosta09} suggested that here may be two modes of star
  formation in dwarf galaxies, a ``normal'' mode with $r_h\sim 3$ pc
  and an ``extended'' mode with $r_h\sim 10$ pc. This ``extended''
  mode, if real, could be potentially related to the excess of large
  GCs we find in FCC~21.}
We do not exclude the GC system of FCC21 from the sample at this time,
but will pay close attention to it in what follows.

\begin{figure*}
\epsscale{1}
\plotone{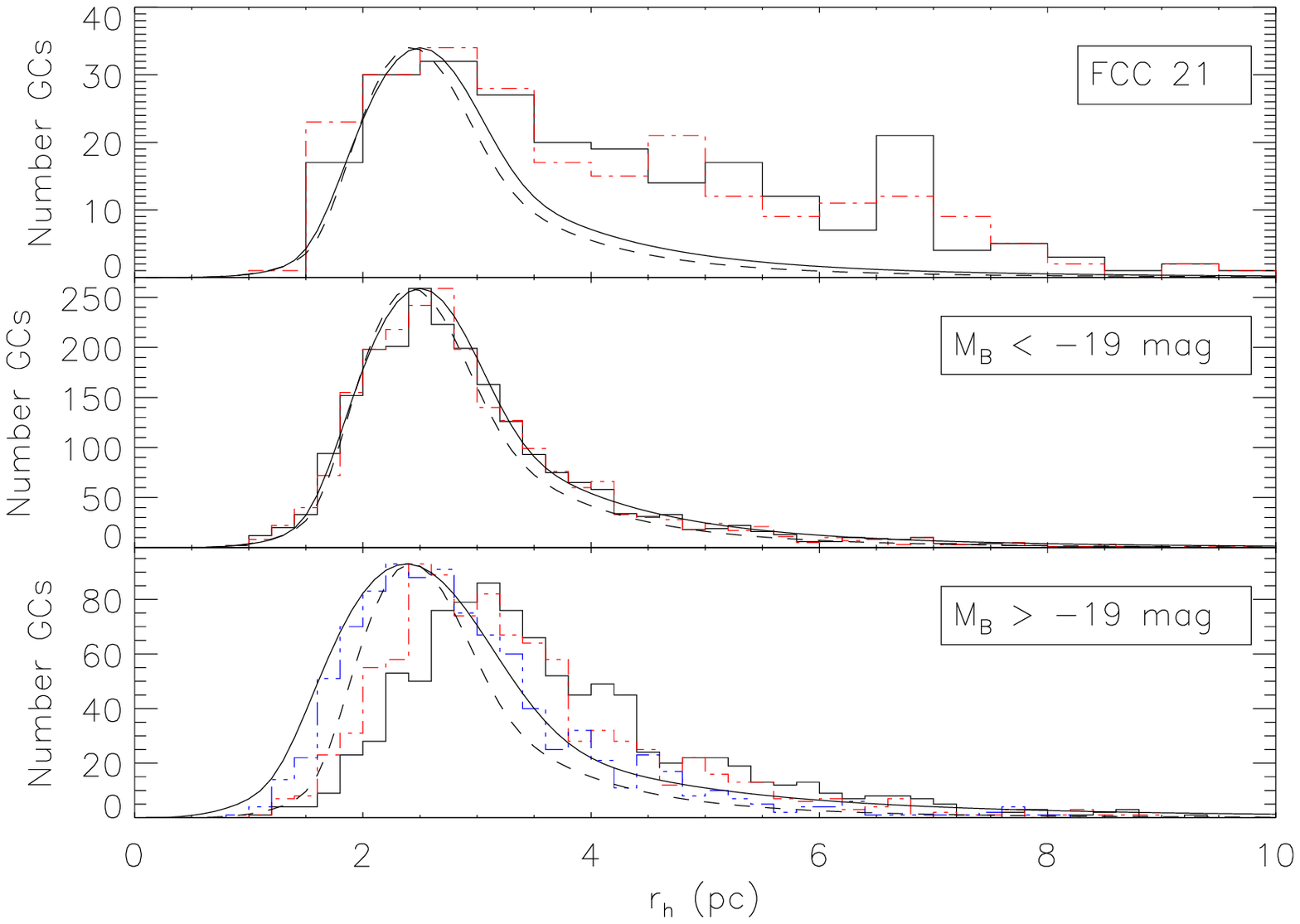}
\caption{The distribution of uncorrected ($r_h$ in black solid
histogram) and corrected ($\hat{r}_h$ in red, dot-dash histogram)
values (in pc) for (1) FCC 21, (2) all other bright Fornax galaxies
($M_B < -19 $mag) and (3) fainter Fornax cluster early types ($M_B >
-19$ mag). Also shown for the fainter galaxies is the ``extra''
corrected, $\hat{r}'_h$ (in blue dot-dot-dot-dash histogram).  The
dashed curve shows the function fitted in J05 to the distribution of
$\hat{r}_h$ from ACSVCS with its peak normalized to the peak of the
observed distribution, while the solid line in the two upper panels
are our best fit of the same function to the bright Fornax cluster
galaxy GC systems, and in the lower panel it is the fit to the fainter
Fornax cluster galaxy GC systems.
\label{rhhists2}}
\end{figure*}

We show in Figure \ref{rhhists2} the grouped GC systems of FCC 21
alone, the remaining bright galaxies ($M_B < -19$ mag, excluding
FCC21), and the dimmer galaxies ($M_B > -19$ mag), with the fitting
function whose parameters are given in Equation 27 of J05 overlayed (dashed curve). The shape of the $r_h$ distribution for the fainter galaxies is
very similar to the luminous galaxies, but the mean is shifted to a
higher value. We also show here the distribution of corrected
$\hat{r}_h$ values (red, dot-dashed histogram) and for the dimmer
galaxies the ``extra'' corrected, $\hat{r}'_h$ values (blue,
dot-dot-dot-dashed histogram) as discussed in Section 7. We perform a
maximum likelihood fit of the function described in Eqns. 24 and 25 of
J05 to the histograms of $\hat{r}_h$ for the bright galaxies and
$\hat{r}'_h$ for the dimmer galaxies finding best fit values of the
parameters ($\tilde{\mu}$, $\tilde{\beta}_1$, $\tilde{\beta}_2$ and
$\tilde{f}$) of (2.65, 0.32, 0.18, 0.70) and (2.64, 0.34, 0.25, 0.61)
within a few $\sigma$ of the quoted dispersion on these values found
from fits to a range of ACSVCS galaxies in J05 (whose errors
describing the range of fits in Virgo cluster galaxies of different
colors would therefore be reasonable estimates here too). These
functional fits are overlayed in Figure \ref{rhhists2} as the solid
curves. It is clear that the fit to the distribution of GC sizes in the
bright Fornax cluster galaxies is very similar to that found for Virgo galaxies by J05,
further demonstrating the universality of the distribution of GC sizes
in early type galaxies, especially the most luminous ones.
 
\section{Factors Which Might Affect GC Size}

It is plausible that many factors may influence the size of a
GC. This could include properties of the GC themselves (such as mass,
metallicity or age) as well as external factors related both to the
local environment at the position of the GC, as well as differences in
the type or size of galaxy in which the GC formed and evolved. Following J05
we separate possible dependencies of the GC half-light radius into three
categories

\begin{itemize}
\item[1.] {\it Internal Factors} in which we consider trends of the
  half-light radius with properties of the GCs themselves. We use the
  $z$-band absolute magnitude (as a proxy for GC mass) and the $(g-z)$
  colors which correlate with GC metallicity and age. Both quantities
  are measured directly from the ACS imaging.
\item[2.] {\it Local Factors} in which we consider trends of $r_h$
  with tracers of the local environment of the GC. We consider the
  GC's position in it's host galaxy, using the galactic radius scaled
  to the effective radius of the host galaxy ($r_p/r_e$) and also the
  local surface brightness, $\mu_z$, and local color $\mu_g-\mu_z$ at
  the position of the GC. The derivation of these quantities from ACS
  images are described in \citet{F06}.
\item[3.] {\it Global Factors} in which we consider trends of $r_h$
  with global properties of the host galaxy. For this we consider the
  absolute B-band magnitudes, $M_B$ (from the RC3, \citet{RC3}), as
  well as the $(g-z)$ color and the average z-band surface brightness
  both measured within the effective radius, $r_e$ from the ACS
  images. The galaxies FCC 167 and FCC 26 have very poor fits to their
  ACS data due to the presence of a massive dust disk and a region of
  star formation the inner region of FCC 167 and FCC 26,
  respectively. In the case of FCC 167 no effective radius could be
  measured (so we take the one in the RC3), and in both cases the
  global colors are difficult to define. For these galaxies we use
  the average color from the areas of the galaxies used for SBF
  measurements in \citet{B09}.
\end{itemize}

As is discussed in J05 there is a significant interdependency between
many of these parameters: for example, it is well know that a galaxy's
total magnitude correlates with color, and the colors of the GCs in a
galaxy are also correlated with both the galaxy's total magnitude and
color.  This makes the problem an ideal candidate for a principal
component analysis (PCA) to look for the primary correlations. Such an
analysis is performed in the Appendix A. We will also
consider the traditional 2D scaling relations below in Section 5.

\section{Trends between Variables}

In this section we consider separately trends between variables, using
traditional 2D plots and constructing best fit relations.  In Appendix
A we describe a PCA which as discussed in Section 4 above is
particularly useful for learning about the shape of the correlations
between many different variables when significant intercorrelations
are expected. However, it is still simpler to interpret the
traditional 2D trends between variables (plotting one as a function of
the other), and by doing this successively with different factors the
interdependencies can also be traced. The PCA in Appendix A has given
us an idea of the shape of the trends we expect to see and which
relations we expect to be important and unimportant (showing that the
variability in $r_h$ can be well described by one factor from each of
the internal, local and global categories we define in Section 4),
however we will consider all possible factors again in what follows.

\subsection{Internal Factors}

In this section we consider how properties of the GCs themselves might
affect the observed sizes of the GCs.

\begin{figure}
\epsscale{1}
\plotone{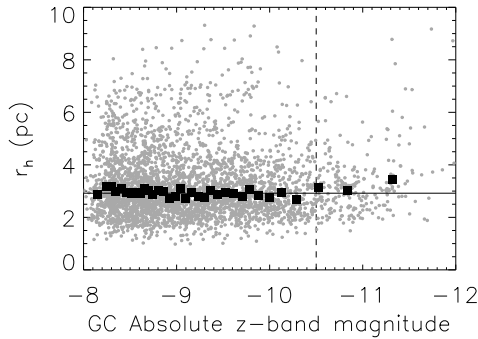}
\caption{Half-light radii ($r_h$) vs. z-band absolute magnitude of GCs
in early type galaxies in the Fornax cluster. Overlayed are the median
values in bins of 100 GCs. The horizontal line shows the median value
for the whole sample of 2.94~pc. We indicate the z-band magnitude
corresponding to $M = 2\times 10^6$ \msun~ (assuming $M/L_z\sim 1.5$
in solar units,
\citealt{J07b}) where a mass-radius relation has previously been shown
to emerge.
\label{avrh_z}}
\end{figure}

It has been seen in previous HST studies (\citealt{KW01}, J05,
\citealt{Barmby07, McLaugh08, Harris2009, H09b, Madrid09}) that the size
  of GCs is roughly independent of their mass and luminosity at least
  for GCs with masses below $M = 2\times 10^6$ \msun
  \citep{H05,M09}. As expected, we confirm this property with GCs in
  Fornax cluster galaxies --- we see no significant correlation of $r_h$
  with the luminosity of the GCs (see Figure \ref{avrh_z}). We
  indicate the z-band magnitude corresponding to $M = 2\times 10^6$
  \msun~ (assuming $M/L_z\sim 1.5$ in solar units, \citealt{J07b}); at brighter
  magnitudes (larger masses) than this there is a hint of an upturn in
  the relation of $r_h$ and magnitude. \citet{S06} suggest that in the
  Sombrero galaxy, an Sa/S0 central galaxy in a small group, there is
  a statistically significant trend of the sizes of the red GCs with
  absolute magnitude. Splitting the GCs in our sample into red and
  blue however we see no significant differences in the trend with
  magnitude. Both subsamples of GCs in early-type Fornax cluster
  galaxy are consistent with having no size-luminosity relation.

\begin{figure}
\epsscale{1}
\plotone{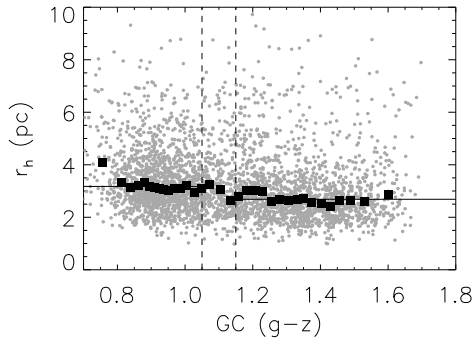}
\caption{Half-light radii ($r_h$) vs. color ($g-z$) of GCs in early
type galaxies in the Fornax cluster. Overlayed are the median values
in bins of 100 GCs. The vertical dashed lines show our color cuts for
blue GCs ($g-z<1.05$) and red GCs ($g-z>1.15$). The horizontal solid
lines show the mean $r_h$ values for the blue and red GCs,
respectively.
\label{avrh_color}}
\end{figure}

\begin{figure}
\epsscale{1}
\plotone{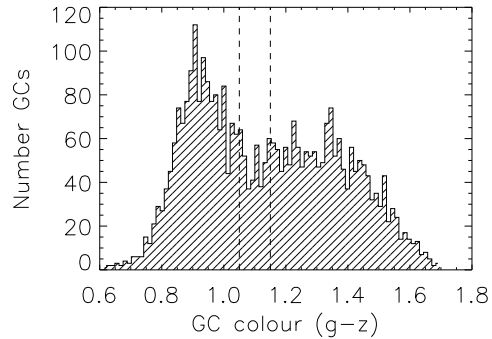}
\caption{Distribution of GC colors for the combined sample of Fornax
cluster galaxy GCs. The vertical solid lines show our color cuts for
blue GCs ($g-z<1.05$) and red GCs ($g-z>1.15$).
\label{GCcolor}}
\end{figure}

As has also been seen in previous HST studies of GCs in bright
ellipticals (\eg~ \citealt{Kundu1999a, Puzia1999a, Larsen2001a}, J05,
\citealt{Madrid09,Harris2009}), we find a clear correlation of $r_h$
with GC color, with bluer GCs having larger half-light radii (see
Figure \ref{avrh_color}). We follow J05 in restricting the sample in
color in order to study other correlations, considering blue GCs
($(g-z) < 1.05$) and red GCs ($(g-z)>1.15$) separately in what
follows, and returning to the issue of the color dependence
later. 
The reason for separating the GCs by color is that many of the
properties of GCs are correlated with GC color, making it hard to
separate the dependencies on each variable. For example, the dependence
of $\langle r_h \rangle$ on GC color will affect the measured
dependence of $\langle r_h \rangle$ with galactocentric radius through
the changing ratio of red to blue GCs with this quantity (see \S3 in
J05 for more examples); by restricting the sample by color we
alleviate these problems.
These color cuts correspond roughly to
cuts in metallicity of [Fe/H]$\lesssim -0.8$ (blue) and [Fe/H]$\gtrsim
-0.65$ (red), where we have used the empirical relation between [Fe/H]
and $(g-z)$ presented in \citet{P06} derived using Milky Way GC data
to transform $(g-z)$ color to [Fe/H].
Within each of these color bins (shown in Figures \ref{avrh_color}
and \ref{GCcolor}) there is little evidence for a trend of $r_h$ with
color. This is especially true for the red GCs --- there may still be a
slight trend for the bluest of the blue GCs to have large values of
$r_h$ than the mean. The median value of $r_h$ for blue GCs is $3.17$
pc, and for red GCs it is $2.69$ pc, using a bi-weight mean we find
$\langle r_h \rangle_{\rm blue} = 3.36\pm 0.03$ pc ($\sigma=1.1$ pc)
and $\langle r_h \rangle_{\rm red} = 2.83\pm 0.02$ pc ($\sigma=1.0$
pc). The systematic error on both these measurements at the distance
of Fornax is $\sim 0.25$ pc (see J05, \S2).

Possible explanations for this observed correlation between $r_h$ and
color are given by projection effects combined with a correlation
between $r_h$ and galactocentric distance similar to that of the Milky Way
\citep{Larsen2003a}, the combined effects of mass segregation and
the dependence of stellar lifetimes on metallicity
\citep{Jordan2004a} or different conditions at formation \citep{Harris2009}.

\subsection{Local Factors}

 \begin{figure*}
\epsscale{0.8}
\plotone{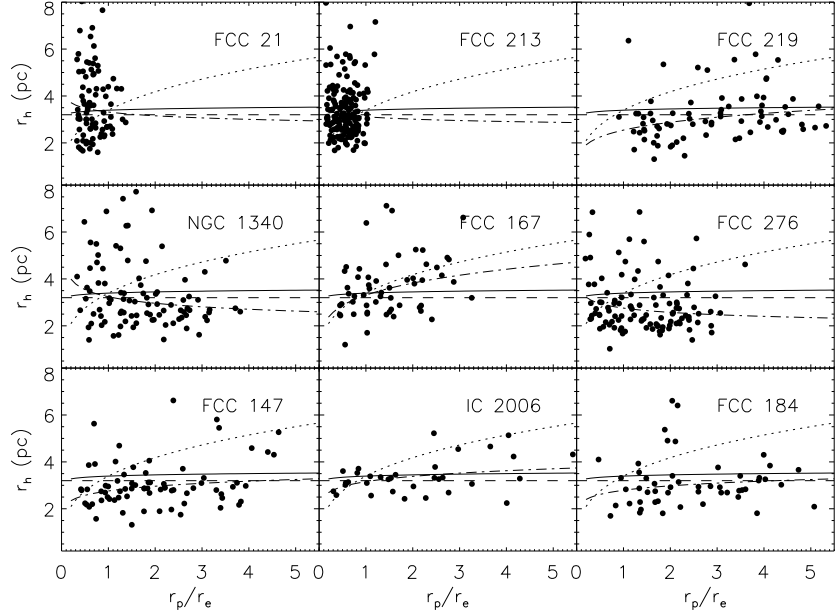}
\caption{GC half-light radius ($r_h$) vs. galactocentric distance (in
  units of the galaxy effective radii) for blue GCs in the nine brightest
  galaxies in the Fornax cluster. The dot-dashed line shows the fits
  to the individual samples done in $\log$-$\log$ space; solid line is
  the weighted mean of the fits to the GC systems in all of the ACSFCS
  galaxies. These can be compared to a zero trend (dashed line -
  plotted at the mean $r_h$ for all blue GCs of $r_h = 3.2$~pc, and the
  relation which would be seen in the MW GC system after projection
  effects are considered (dotted line).
\label{localfactors_r}}
\end{figure*}
 
Here we will look at how the local environment of a GC at the position
it is found in its galaxy might affect the observed size of the
GC. Using the sample of GCs separated into a blue subsample and a red
subsample we now look for correlations with projected galactocentric
radius, $r_p$, local galaxy surface brightness, $\mu_z$ and local
galaxy color, $\mu_g-\mu_z$. Obviously these factors are strongly
correlated. In particular we expect $\mu \propto \log [ F (r_p)]$
where $F$ is some function describing the shape of the surface
brightness profile of a given galaxy. This does not mean that
descriptions in the two different variables are equivalent since the
function $F$ can be quite complicated, however there will be
significant correlation between the two. In ellipticals, it is also
generally the case that there is a color gradient with a bluer mean
color in the outskirts of the galaxy than is observed in the central
regions (as seen in ACSVCS data by \citealt{F06}). Ultimately we will
follow J05 in choosing $\mu_z$ as the most reliable independent
variable, however, to begin with we will consider correlations with all
three factors.

\begin{figure*}
\epsscale{0.8}
\plotone{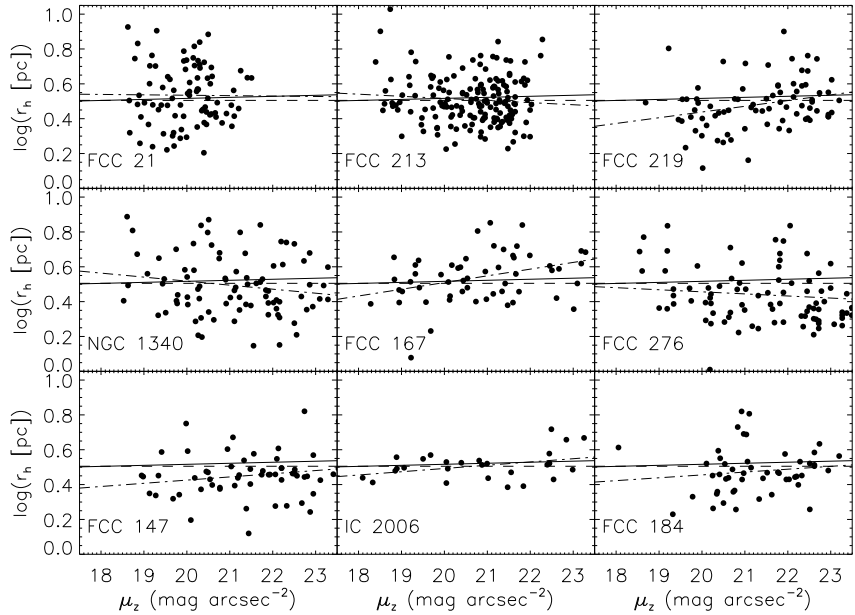}
\caption{GC half-light radius [$\log(r_h)$] vs. local galaxy surface
brightness for blue GCs in the nine brightest galaxies in the Fornax
cluster. The dot-dashed line shows the fits to the individual samples,
solid line is the weighted mean of all the fits to the GC systems in
all of the ACSFCS galaxies. The dashed line shows a zero trend at the
mean of the value of $r_h$ for all blue GCs (3.2 pc).
\label{localfactors_mu}}
\end{figure*}

In order to place all GC systems on the same scale, we normalize their
galactocentric radii using the effective radii of the galaxy, $r_e$ as
measured from the ACS images (Ferrarese et al 2010, in preparation;
note than in FCC 167 heavy dust obscuration prevented fitting the ACS
image --- in this case $r_e$ is taken from the RC3). Figure
\ref{localfactors_r} shows half-light radius, $r_h$ vs. projected
galactocentric distance, $r_p/r_e$ for the blue GCs in the nine brightest
early-type Fornax cluster galaxies. In the two brightest ACSFCS
galaxies (FCC 21 and FCC 213) GCs are only traced to a couple of
effective radii, but the vast majority of ACSFCS galaxies have GC data
out to at least $5r_e$, with the maximum value in the sample being
14$r_e$ (for FCC 249, $r_e=7.6\arcsec$).
 
A small trend is visible in most galaxies, with GCs at larger
galactocentric radii having, on average, slightly larger half-light
radii. We fit simple linear relations of the form $\log(r_h) = a_r + b_r \log
(r_p/r_e)$ to the data in all Fornax cluster galaxies, then take a
weighted average of these results finding $b_r=0.029\pm0.016$. There
is no significant trend of the slope of the relation with the extent
of the observed GC system ($r_{p,max}/r_e$) suggesting that we are not
hiding a stronger relation by being dominated by GCs in bright
galaxies for which the radial extent of observed GCs is small.  In
contrast to the findings of \citet{S06} for the Sombrero galaxy, we do not find a significantly stronger
trend with the red GC population than the blue ones, measuring $b_{r,
{\rm red}} = 0.041\pm0.026$, which is the same as the trend of the blue GCs
to within the $1\sigma$ errors. We do the same fits for $\log r_h$ vs
$\mu_z$ (see Figure \ref{localfactors_mu}) finding in this case
$b_\mu=0.006\pm0.003$ from a weighted mean of the fits for each
galaxy. These results are similar to, but a bit flatter (less
significant) than the relations found in J05 for GCs in Virgo cluster
galaxies of $b_r=0.07\pm0.01$ and $b_\mu=0.016\pm0.003$. We do this
same fit again for the Virgo sample, and find $b_\mu=0.011\pm0.003$
(the slight difference being due to slight changes in the sample and
fitting methods). A combination of GC systems in Fornax and Virgo
galaxies gives $b_\mu=0.008\pm0.002$.
 
\begin{figure*}
\epsscale{0.8}
\plotone{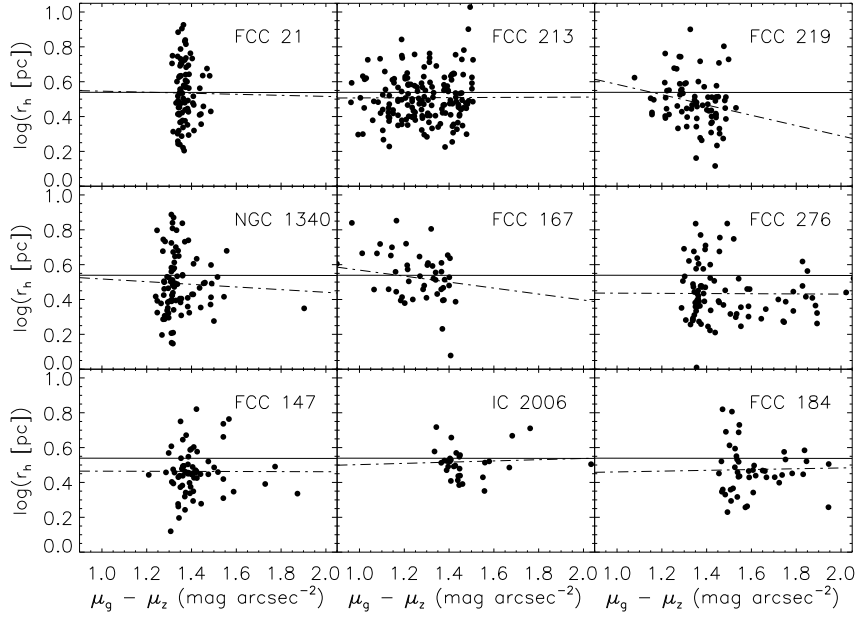}
\caption{GC half-light radius [$\log(r_h)$] vs. local galaxy color
($\mu_g-\mu_z$) for blue GCs in the nine brightest galaxies in the Fornax
cluster. The dot-dashed line shows the fits to the individual samples,
solid line is the weighted mean of all the fits to the GC systems in
all of the ACSFCS galaxies. The dashed line shows a zero trend at the
mean of the value of $r_h$ for all blue GCs (3.2 pc).
\label{localcolor}}
\end{figure*}

We also consider the radii of the blue GC vs. the local color of
the galaxy ($\mu_g - \mu_z$), which is shown in Figure
\ref{localcolor}. The outer parts of the early-type galaxies in
Fornax are all bluer than the inner parts. This corresponds to a trend
with GC half-light radius such that those with redder local galaxy
colors are, on average, slightly smaller. However, a fit of the form
$\log(r_h) = a_c + b_c (\mu_g - \mu_z)$ results in a mean value of
$b_c = -0.012\pm0.013$ --- a negligible trend.

These results come from the blue GCs only, but no significant
differences are seen if just the red GCs are considered. The exclusion
of GCs in the peculiar GC system of FCC 21 (= NGC 1316, Fornax A)
which are currently included, also has little impact on the final
results here.

We now correct the half-light radii of the GCs to what is expected for
a GC with an underlying surface brightness of 21 mag
arcsec$^{-2}$. Since we want a global correction for all GCs we will
use the slope of trend fit to both Fornax and Virgo cluster early-type
GC systems together of \be r'_h \equiv r_h 10^{-0.008(\mu_z -21)}.
\ee There is evidence however that the trend is slightly stronger in
the Virgo galaxies, and slightly weaker (possibly zero) in the Fornax
galaxies. This is also a weaker trend than in J05, who used
$r'_h \equiv r_h 10^{-0.016(\mu_z -21)}$.

\subsection{Global Factors}
We now look for correlations of the GC half-light radius with global
properties of the host galaxy, such as total $B$-band luminosity and
the average surface brightness and ($g-z$) color within the effective
radius. We use the half-light radius, $r'_h$ corrected for local
effects, and consider the blue ($(g-z)_{GC} < 1.05$) and red
($(g-z)_{GC} > 1.15$) GCs separately.
 
\begin{figure}
\epsscale{1}
\plotone{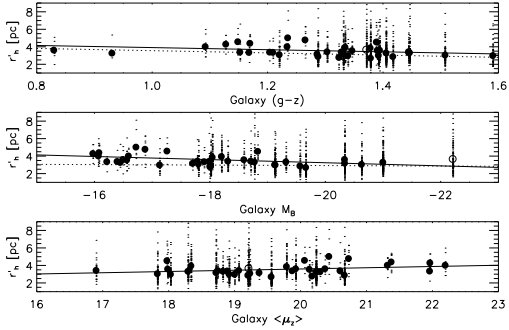}
\caption{Top: $r'_h$ vs galaxy color for blue GCs. The filled circles
show the biweight mean of the distribution for each galaxy which has three
or more blue GCs. The solid line is fit to these averages. The open
circle is the outlier FCC21 which is not included in the fits. Dotted
lines show the fits from J05 to Virgo cluster galaxies. Middle panel:
the same, but for $r'_h$ vs galaxy total B-band magnitude. Bottom: the
same for $r'_h$ vs galaxy average surface brightness.
\label{globalfactors}}
\end{figure}

Figure \ref{globalfactors} shows the result for blue GCs in our sample
of early type Fornax cluster galaxies. All blue GCs are shown; the
solid circles show $\langle r'_h \rangle $ (biweight mean of $r'_h$)
for each galaxy which has three or more blue GCs. We exclude FCC 21 (NGC
1316) from the fits as it is much more luminous than all other
galaxies and has a peculiar GC size distribution (as shown in Section
3).  The dotted line shows the trends with galaxy color and blue
magnitude observed by J05. We find\footnote{Note that we are using a
short-hand notation for errors here where $0.515(20) = 0.515\pm
0.020$.}
\begin{eqnarray}
\log  \langle r'_{h,{\rm blue,Fornax}} \rangle & = & 0.515(20) - 0.142(75)[(g-z)_{\rm gal} - 1.5] \nonumber \\
\log  \langle r'_{h,{\rm blue,Fornax}} \rangle & = & 0.500(18) + 0.023(07)(M_B + 20) \nonumber \\
\log  \langle r'_{h,{\rm blue,Fornax}} \rangle & = & 0.534(12) + 0.018(09)(\langle\mu_z\rangle -19).
\end{eqnarray}
We refit the same relations to Virgo cluster galaxies  --- using a sample that is similar, but not identical, to that used in J05 (as discussed in \S2) --- and find:
\begin{eqnarray}
\log  \langle r'_{h,{\rm blue,Virgo}} \rangle & = & 0.453(10) - 0.225(45)[(g-z)_{\rm gal} - 1.5] \nonumber \\
\log  \langle r'_{h,{\rm blue,Virgo}} \rangle & = & 0.468(11) + 0.014(04)(M_B + 20)
\end{eqnarray}
Note that as well as this sample being slightly different to that used
in J05 (possibly of particular importance here, some low-luminosity
galaxies are added), the correction for local effects (a correlation
of $r_h$ with local surface brightness) is also slightly shallower
than what was used in J05. These difference presumably explain the
slight changes seen in the relationships we find and those reported by
J05.

The slope of the $\log \langle r'_{h,{\rm blue}} \rangle$ vs. galaxy
color for Fornax galaxy GC systems is very slightly shallower than
that observed for Virgo cluster galaxies (the same within $2\sigma$);
both are consistent with the $-0.167\pm0.054$ found by J05.  We find a
trend with magnitude at the $\sim 3 \sigma$ level in both the Fornax
cluster and Virgo cluster galaxies (both the trend and the error on it
are twice as large in Fornax than in Virgo), while the slope found for
Virgo by J05 was consistent with zero within the 1$\sigma$ errors. We
attribute this difference to the slightly different samples
and corrections for local effects as described above. J05 did
not explore trends with the average surface brightness of the galaxy
within the effective radius. In any case, we find this trend to be very
small --- only different from zero at about the $2\sigma$ level.

Repeating this analysis using red GCs in Fornax/Virgo cluster galaxies,
we find slight differences such that the trends of size with color
and magnitude for red GCs are slightly larger than for blue GCs. We
still find that the trend with color is (slightly) larger in the
Virgo cluster galaxies, while the trend with magnitude is larger in
the Fornax cluster galaxies. The trend with average surface brightness
is still consistent with zero (in fact, the sign flips). We also repeat
the fits only including GCs within $5r_e$ in the Fornax cluster
galaxies (and removing FCC 213 along which FCC 21 which was already
not included) to make sure that these trends are not biased by the
fact that the fainter, bluer galaxies typically have GCs to larger
radial distances than the more massive ones. This cut does not
significantly change the fits.

These differences with the trends of half-light radius with global
galaxy properties in Fornax early types vs. Virgo early types and also
with blue and red GCs suggests that care should be taken applying
these corrections, and further study is needed to determine if there
truly is a universal correlation. They may also point to interesting
differences in the formation and evolution of GCs in Fornax cluster
galaxies versus Virgo cluster galaxies and provide more constraints on
the differences between the blue and red subpopulations of GCs. It is
possible that different selection procedures for the dwarf galaxies
included in the ACSVCS versus the ACSFCS could cause the differences
between Virgo and Fornax. However, the impact of selection effects on
the galaxies on trends of the average sizes of the GCs with global
galaxy properties is likely to be complicated.
 
\begin{figure}
\epsscale{1}
\plotone{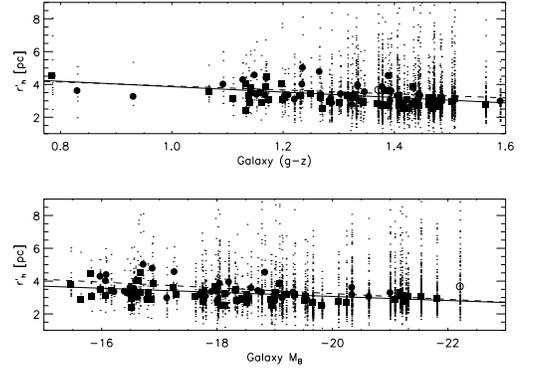}
\caption{As Figure \ref{globalfactors} but for blue GCs in both Fornax
and Virgo cluster galaxies; biweight means for Fornax galaxies are
shown as circles, for Virgo they are squares. The solid line are the
fits to this data, dashed lines show the equivalent fits for just the
blue GCs in Fornax galaxies (as in Figure \ref{globalfactors}).
\label{globalall}}
\end{figure}

 Figure \ref{globalall} shows the same plot as in Figure 9, but for
 blue GCs in both Virgo and Fornax cluster galaxies combined. The fits
 in this case are

\begin{eqnarray}
\log  \langle r'_{h,{\rm blue,both}} \rangle & = & 0.473(10) - 0.209(41)[(g-z)_{\rm gal} - 1.5] \nonumber \\
\log  \langle r'_{h,{\rm red,~both}} \rangle & = & 0.409(10) - 0.575(91)[(g-z)_{\rm gal} - 1.5] \nonumber \\
\log  \langle r'_{h,{\rm blue,both}} \rangle & = & 0.480(10) + 0.018(04)(M_B + 20)\nonumber \\
\log  \langle r'_{h,{\rm red,~both}} \rangle & = & 0.434(14) + 0.039(07)(M_B + 20).
\end{eqnarray}
The addition of the systems in Virgo slightly reduces the significance
of the trend with galaxy magnitude, and there is a clear difference in
the average $r'_h$ values between Virgo and Fornax for galaxies fainter
than --18 mag (with Fornax GC systems having larger average $r'_h$ than
Virgo). The slope of the trend with galaxy color changes mostly because
of the offset in zeropoint between Fornax and Virgo, combined with the
distribution of colors of the galaxies in the two clusters (i.e., the
slope for Virgo is parallel to but offset from that in Fornax).

\begin{figure}
\epsscale{1} 
\plotone{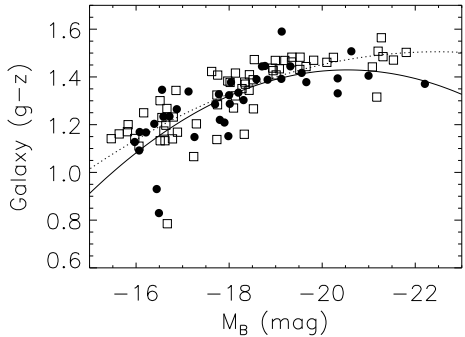}
\caption{Color-magnitude diagram for all galaxies in the ACSVCS and ACSFCS. Fornax galaxies are shown as filled circles, while Virgo are open squares. The relation fit by J05 is shown as the dotted line. A quadratic fit to the Fornax data is shown by the solid line. The red outlier at $M_B \sim -19$ mag is FCC184, a face-on SB0 which may suffer from significant internal reddening.
\label{colormagnitude}}
\end{figure}

Following J05 we argue that galaxy color and $M_B$ should be almost
equivalent as tracers of the global variation in $r'_h$. We show the
color magnitude diagram for both ACSVCS and ACSFCS galaxies in Figure
\ref{colormagnitude}. In J05 it was shown that folding the linear
variation of $\log r'_h$ with a quadratic relation describing the
correlation between galaxy color and luminosity in the ACSVCS
galaxies reproduces well the mild dependence of $r'_h$ on $M_B$. We
find indications of a steeper trend of average $r'_h$ with galaxy
magnitude in Fornax, and a shallower trend with galaxy color, even
though there is no significant difference in the color-magnitude
diagram. The biggest difference in the Virgo and Fornax color-magnitude diagrams
shown in Figure \ref{colormagnitude} is that in the Fornax cluster
the most luminous galaxies are not as red on average as they are in
the Virgo cluster (Ferrarese et~al. 2006, 2010). It is perhaps this slight blueward compression of
the colors of Fornax early types which means that the color trend
alone cannot explain all the dependence of GC size on galaxy
mass. Among the Fornax cluster galaxies, the red outlier at $M_B \sim
-19$ mag is FCC184, a face-on SB0 which may suffer from significant
internal reddening. FCC 167 and FCC 26 (two galaxies with poor fits to
the ACS images for which we use instead of the average color within
$r_e$, the average color of regions used for the SBF measurement in
\citet{B09}) are the bluest points at $M_B\sim -20$ and $-16.5$
respectively.

In order to fix the correlation of GC sizes with the mass of the
galaxies, a trend with the galaxy magnitude and not just the color is
needed, especially in low-luminosity Fornax cluster galaxies. However
the correlation between $\log r'_h$ and galaxy $(g-z)$ color that we
find in both Fornax and Virgo cluster galaxy GC systems is consistent
with J05, so we adopt the same correction to $r''_h$ used there, 
\be r''_h \equiv r'_h 10^{0.17[(g-z)_{\rm gal}-1.5]},  \ee or, combining
with the correction for local environment (to $r'_h$) \be
r''_h \equiv r_h 10^{-0.008(\mu_z-21)+0.17[(g-z)_{\rm gal}-1.5]}, \ee

We will return to the issue of a possible extra dependence of average $r'_h$ value with magnitude in \S8. 

\subsection{Dependence on GC Color}

We now return to the issue of dependence of GC half-light radius on
color, $(g-z)_{\rm GC}$. We use the half-light radius corrected for
both local and global factors as described above,~$r''_h$.

\begin{figure}
\epsscale{1}
\plotone{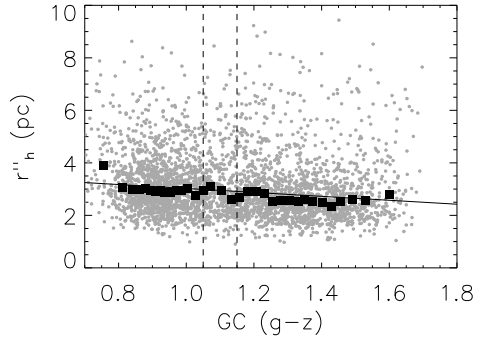}
\caption{Corrected half-light radius, $r''_h$ vs. GC color for all
GCs in all of our sample galaxies combined. The large squares show
averages (biweight means) in bins of 0.15 mag of color. The solid
line shows the best fit linear relation of the form: $\log \langle
r''_h \rangle \propto (g-z)_{\rm GC}$.
\label{gccolor}}
\end{figure}

Figure \ref{gccolor} shows the corrected half-light radius, $r''_h$
vs. GC color for all GCs in all of our sample galaxies combined. The
large squares show averages (biweight means) in bins of 0.15 mag of
color. The solid line shows are best fit to a linear relation of the
form $\log \langle r''_h \rangle \propto (g-z)_{\rm GC}$, the best fit
slope is $b=-0.12\pm0.01$. This is similar to the relation found for GCs in
Virgo galaxies by J05, who found $\log \langle r''_h \rangle \propto
-(0.17 \pm 0.02)(g-z)_{\rm GC}$) but used a slightly different
correction for local factors, as discussed above.

J05 contain a lengthy discussion of the correlation of GC color with
galaxy color and luminosity. Less luminous, bluer ellipticals have GC
systems which are usually metal poor, so have very few red GCs and
therefore add little leverage to any trends of the GC sizes with GC
color which will be dominated by the trends in more massive
galaxies. In order to test that the correlation between half-light
radius and GC color is the same in all colors and luminosities of
galaxies we follow J05 in defining subsamples of GCs in galaxies of
similar colors. We exclude any galaxy which has extremely small
numbers of very red GCs (they must have a number of GCs with colors
$(g-z)_{\rm GC} > 1.3$ which is equal to at least 10\% of the blue GCs
$(g-z)_{\rm GC} < 1.05$) since these galaxies add little or no
information about the color dependence of the half-light radius. We
then create bins in galaxy color, by requiring at least 80 very red
GCs ($(g-z)_{\rm GC} > 1.3$) in each bin; some bins have only a single
galaxy. 

\begin{figure*}
\epsscale{1}
\plotone{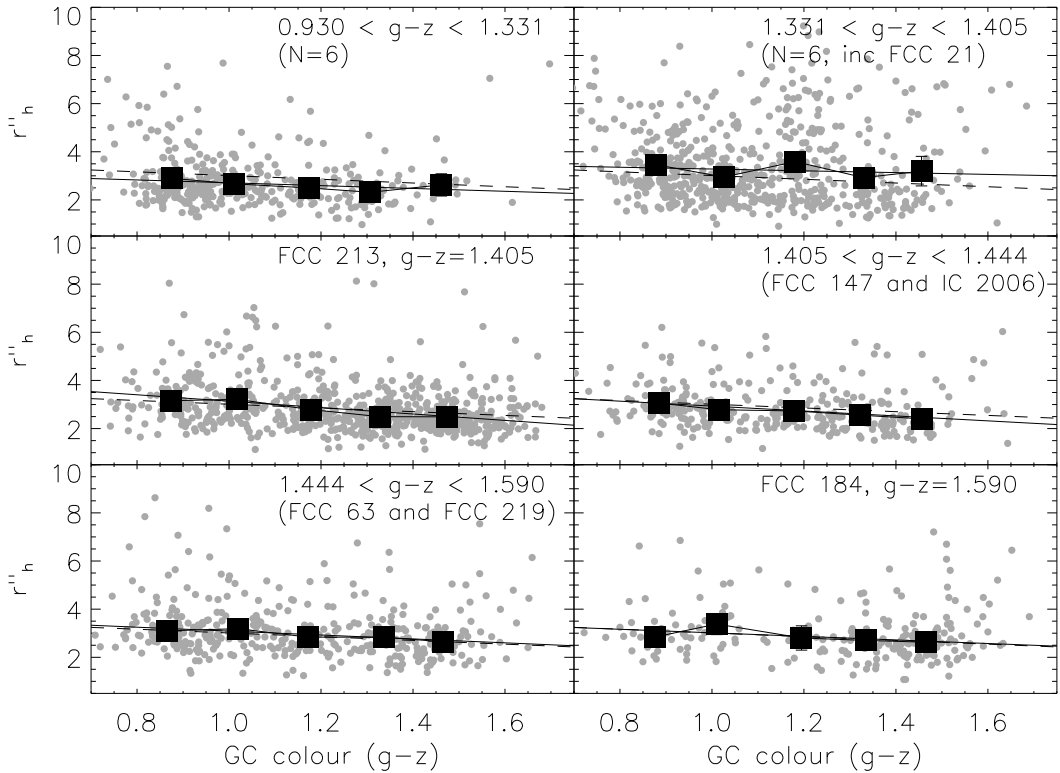}
\caption{As Figure \ref{gccolor} but in subsamples separated by
galaxy color (as indicated). The dashed line in each panel shows the
fit to the entire sample in Figure \ref{gccolor}.
\label{gccolor_bins}}
\end{figure*}

The results of this exercise are shown in Figure
\ref{gccolor_bins}. All subsamples are consistent with having $\log
\langle r''_h \rangle \propto -(0.12-0.17) (g-z)_{\rm
  GC}$. Interestingly, there is quite a wide range of values of the
extent of the galactocentric radii of the GCs ($r_p/r_e$) in these
subsets, which vary from $r_p/r_e = $ 0.1--1.2 (in the bin containing
FCC 213) to $r_p/r_e = $ 0.5--8.7 (in the bin containing FCC 63 and FCC
219). This argues against projection effects \citep{Larsen2003a} being
the primary mechanism explaining the $r_h$--color correlations. Beyond
the large ACSVCS and ACSFCS samples, other studies of the dependence of
$r_h$ with GC color also suggest that this dependence is at least
partly intrinsic to the clusters (see \S3.2 of \citealt{H09b} for a
summary of this evidence; note that an earlier claim by \citealt{S06}
that projection effects are responsible for the $r_h$--color
correlation in the Sombrero galaxy has been revised in
\citealt{H09b}). Possible mechanisms for creating this intrinsic
dependence of size and color have been given in \citet{Jordan2004a}
and \citet{Harris2009}.

\section{Final Correlations of Half-Light Radius}

In this section we list the final correlations of the half-light
radius of the GCs in Fornax cluster galaxies with internal, local and
global environmental variables (as listed in \S4). We find
subtly different scaling relations for the size of GCs in Fornax
cluster galaxies than were found by J05 in Virgo cluster galaxies. We find that
the trend with local environment (measured by local surface
brightness) may be a bit flatter in the Fornax cluster galaxies than
in Virgo cluster galaxies ($b_{\mu, {\rm Fornax}} = 0.006\pm0.003$ as
opposed to $b_{\mu, {\rm Virgo}} = 0.011\pm 0.003$, while the combined
sample gives $b_{\mu} = 0.008\pm 0.002$).  After applying this local
correction, we look at correlations with global properties of the
galaxies and find a similar trend of average size with galaxy color
as was found in the GC systems of Virgo cluster galaxies, but there is some evidence
that an extra correlation with galaxy magnitude might be present in the GC systems of
Fornax dwarfs. After applying both corrections for local and global
properties finally we find a slightly shallower trend of GC size with
GC color ($\log \langle r''_h \rangle \propto -0.12 (g-z)_{\rm GC}$
vs $\log \langle r''_h \rangle \propto -0.17 (g-z)_{\rm GC}$ in Virgo cluster
galaxy GCs).

We therefore use corrected radii
\be
\hat{r}_h \equiv r_h 10^{-0.008(\mu_z-21)+0.17[(g-z)_{\rm gal}-1.5]+0.12[(g-z)_{\rm GC}-1.2]}.
\ee
We also consider an extra correction of 
\be
\hat {r'}_h \equiv \hat {r_h} 10^{-0.02[M_B+21]}
\ee
which is only important for galaxies with $M_B \gtrsim -19$ mag in the Fornax cluster. 

These subtle differences in the trends raise clear questions about the
applicability of using the average half-light radii of GCs belonging
to low-luminosity galaxies as a distance
indicator. On the other hand, this may be hinting at an interesting difference between
the formation/evolution of GC systems in dwarf galaxies in the
Virgo cluster versus the Fornax cluster --- possibly as a result of the
different environments.

\begin{figure*}
\epsscale{1}
\plotone{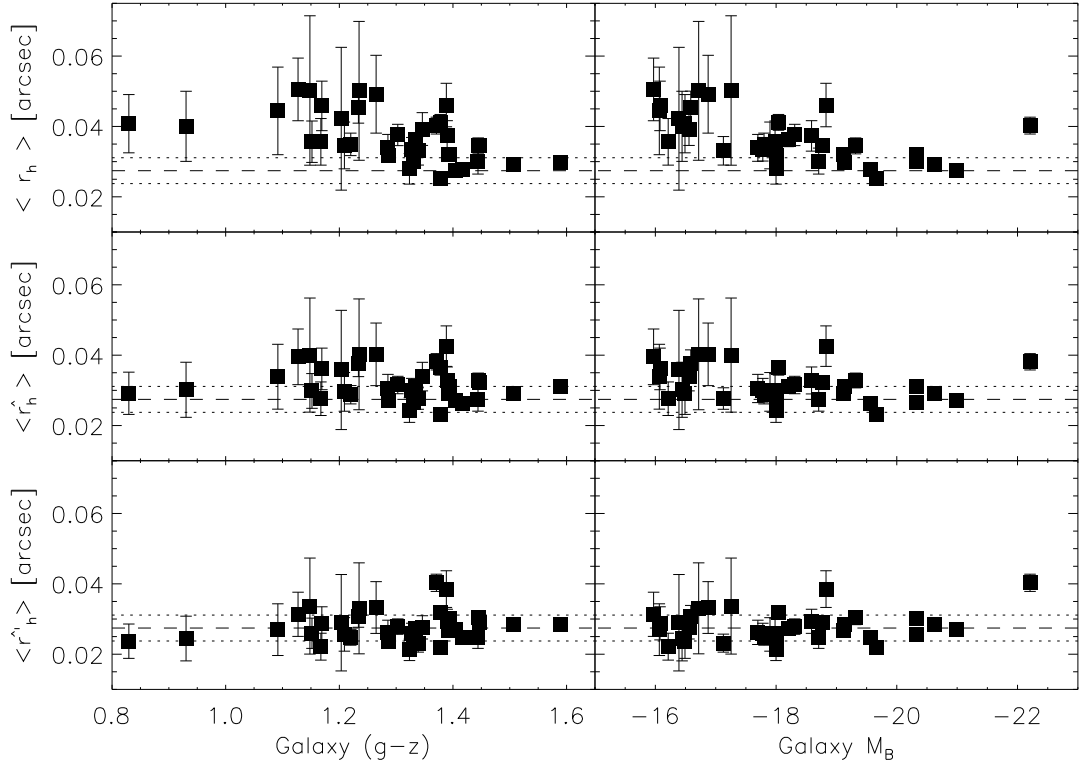}
\caption{Corrected, $\hat{r}_h$ (middle panel; corrections for galaxy
and GC color and local disk surface brightness), extra corrected
(lower panel - including a galaxy magnitude term) and uncorrected,
$r_h$ (top panel) vs. galaxy color and magnitude for all galaxies in
the sample. The dashed lines show the biweight mean of the extra
corrected values excluding FCC 21 (dashed line show the sigma, also
calculated using a biweight location estimator). Error bars on the
points are the 95\% confidence internals on the means for each
galaxy. \label{correctedrh}}
\end{figure*}

We provide in Table~\ref{avrh} a listing of the ACSFCS galaxies 
for which we have $r_h$ measurements. Galaxies are ordered by absolute B-band magnitude.
Also included in this table are the measured
color of the galaxy, average surface brightness at
the position of the GCs, average color of the GCs and the average
sizes of the GCs (raw, corrected and ``extra corrected"). Figure
\ref{correctedrh} shows the corrected, $\langle\hat{r}_h\rangle$,
extra corrected $\langle\hat {r'}_h\rangle$, and uncorrected, $\langle
r_h\rangle$ vs. galaxy color and magnitude for all galaxies in the
Fornax sample. The dashed lines show the biweight mean of the extra
corrected values excluding FCC 21 (dashed line show the sigma, also
calculated using a biweight location estimator). Error bars on the
points are the 95\% confidence intervals on the means for each
galaxy. It is clear that after the corrections are applied the value
of $\langle\hat{r}_h\rangle$ is remarkably consistent over galaxy
color and magnitude. Nevertheless, two outliers are obvious, having larger than usual
values of $\langle\hat{r}_h\rangle$. The first (at $M_B \sim -22$) is
FCC 21 (NGC 1316 or Fornax A), which was noted earlier as a
merger remnant showing a 
peculiar distribution of GC half-light radii. The second (at $M_B \sim -19$) is FCC193,
an S0 galaxy close to the centre of the Fornax cluster.  Unlike J05 we
do not find that the average $r_h$ is consistent with being constant
after just corrections for galaxy and GC color and local surface
brightness of the disk. We find that, in Fornax dwarfs, the average
value of $r_h$ is larger than expected (as is apparent from the middle panel
of Figure~\ref{correctedrh}).
 
\begin{deluxetable*}{lcccccccc}
\tablewidth{0pc} 
\tablecaption{Summary of GC sizes in Fornax Cluster Galaxies}
\tablehead{\colhead{Galaxy} & \colhead{$M_B$} &  \colhead{$(g-z)_{Gal}$} &  \colhead{$\langle \mu_z(r_p) \rangle$} & \colhead{$N_{\rm GC}$} &  \colhead{$\langle (g-z)_{GC}\rangle$} &  \colhead{$\langle r_h \rangle (\arcsec) $}&  \colhead{$\langle \hat{r}_h \rangle (\arcsec) $}  & \colhead{$ \langle \hat{r}'_h \rangle (\arcsec)$} \label{avrh}}
\startdata 
FCC 21&  -22.21& 1.37& 20.07& 1.12& 232 & 0.040(02)& 0.038(02)& 0.040(02)\\
FCC 213& -21.00& 1.41& 20.63& 1.26&  698 & 0.028(01)& 0.027(01)& 0.027(01)\\
FCC 219&  -20.63& 1.51& 21.08& 1.19& 220 & 0.029(01)& 0.029(01)& 0.029(01)\\
NGC 1340&  -20.33& 1.33& 21.14& 1.00& 137 &0.030(02)& 0.027(02)& 0.026(02)\\
FCC 167&  -20.33& 1.09& 20.69& 1.24& 266 &0.032(01)& 0.031(01)& 0.030(01)\\
FCC 276&  -19.66& 1.38& 21.51& 1.11& 232 & 0.025(01)& 0.023(01)& 0.022(01)\\
FCC 147&  -19.56& 1.42& 21.76& 1.15&  201 &0.028(01)& 0.026(01)& 0.025(01)\\
IC 2006&  -19.32& 1.44& 21.31& 1.15&  89 &0.034(02)& 0.033(02)& 0.030(02)\\
FCC 184&  -19.13& 1.59& 21.44& 1.29& 216 & 0.030(01)& 0.031(01)& 0.028(01)\\
 FCC 83&  -19.12& 1.39& 21.55& 1.08& 173 & 0.032(01)& 0.029(01)& 0.027(01)\\
\hline
FCC 193&  -18.83& 1.39& 21.09& 1.08&  23 & 0.046(06)& 0.043(06)& 0.039(05)\\
 FCC 63&  -18.77& 1.45& 21.91& 1.07& 142 & 0.034(02)& 0.032(02)& 0.029(01)\\
FCC 170&  -18.71& 1.44& 21.89& 0.98& 30 & 0.030(04)& 0.027(03)& 0.025(03)\\
FCC 153&  -18.59& 1.39& 21.34& 0.92& 28 & 0.037(04)& 0.033(04)& 0.029(03)\\
FCC 177&  -18.31& 1.30& 22.82& 0.96&  47 &0.038(03)& 0.032(02)& 0.028(02)\\
FCC 249&  -18.20& 1.33& 22.65& 1.00& 102 &0.036(02)& 0.031(02)& 0.027(02)\\
FCC 190&  -18.04& 1.38& 22.08& 0.98&  103 &0.041(02)& 0.036(02)& 0.032(02)\\
 FCC 47&  -18.01& 1.29& 22.62& 1.05& 184 &0.032(01)& 0.027(01)& 0.024(01)\\
FCC 310&  -18.00& 1.32& 20.64& 0.93&  18 & 0.028(04)& 0.024(04)& 0.021(03)\\
 FCC 43&  -17.98& 1.15& 20.81& 1.03& 15 & 0.036(06)& 0.030(05)& 0.026(04)\\
FCC 148&  -17.90& 1.21& 21.41& 1.02& 13 & 0.035(07)& 0.030(05)& 0.026(05)\\
FCC 255&  -17.80& 1.22& 22.56& 0.99& 57 & 0.035(03)& 0.029(03)& 0.025(02)\\
FCC 277&  -17.78& 1.33& 21.61& 1.01& 17 & 0.034(03)& 0.030(03)& 0.026(03)\\
 FCC 55&  -17.70& 1.29& 20.49& 1.08& 12 & 0.034(04)& 0.030(04)& 0.026(03)\\
FCC 152&  -17.25& 1.15& 21.04& 0.86& 5 &  0.050(21)& 0.040(16)& 0.034(14)\\
FCC 143&  -17.13& 1.34& 22.83& 0.94& 33 & 0.033(04)& 0.028(03)& 0.023(03)\\
 FCC 95&  -16.88& 1.26& 24.51&  0.99&13 & 0.049(11)& 0.040(09)& 0.033(07)\\
FCC 204&  -16.72& 1.23& 22.08& 0.85& 7 & 0.050(20)& 0.040(16)& 0.033(13)\\
FCC 136&  -16.59& 1.23& 22.62& 0.97& 13 & 0.045(05)& 0.038(04)& 0.031(03)\\
FCC 182&  -16.56& 1.35& 23.29& 1.02&22 &  0.039(05)& 0.034(04)& 0.028(03)\\
 FCC 26&  -16.49& 0.64& 22.03& 1.01& 14 & 0.041(08)& 0.029(06)& 0.024(05)\\
 FCC 90&  -16.44& 0.93& 21.71& 0.99& 8 & 0.040(10)& 0.030(08)& 0.024(06)\\
FCC 106&  -16.39& 1.20& 21.43& 1.07& 5 & 0.042(20)& 0.036(17)& 0.029(14)\\
FCC 324&  -16.21& 1.17& 22.80& 0.89& 7 & 0.036(07)& 0.028(05)& 0.022(04)\\
FCC 203&  -16.08& 1.17& 22.76& 0.91& 10 & 0.046(07)& 0.036(06)& 0.029(05)\\
FCC 100& -16.07& 1.09& 23.10& 0.94& 9 & 0.044(12)& 0.034(09)& 0.027(07)\\
FCC 303& -15.97& 1.13& 22.89& 0.95& 10 & 0.051(09)& 0.040(08)& 0.031(06)
\enddata
\end{deluxetable*}

\section{Discussion}

\subsection{Half-Light Radius as a Distance Indicator}

Despite the differences in the environmental trends of GC half-light
radius examined here, we  begin this section by re-emphasizing
how little the average sizes of GCs actually vary from galaxy to
galaxy. The observed trends are all very mild and result in only small
changes in the average half-light radii over significant changes in
host galaxy color and luminosity. This is particularly true in the
case of the brightest early-type galaxies in which the uncorrected
mean half-light radius of the GC system already offer a distance
indicator with errors comparable to those from the best techniques currently available (i.e., $\sim$
10-15\%).

\begin{figure*}
\epsscale{1}
\plotone{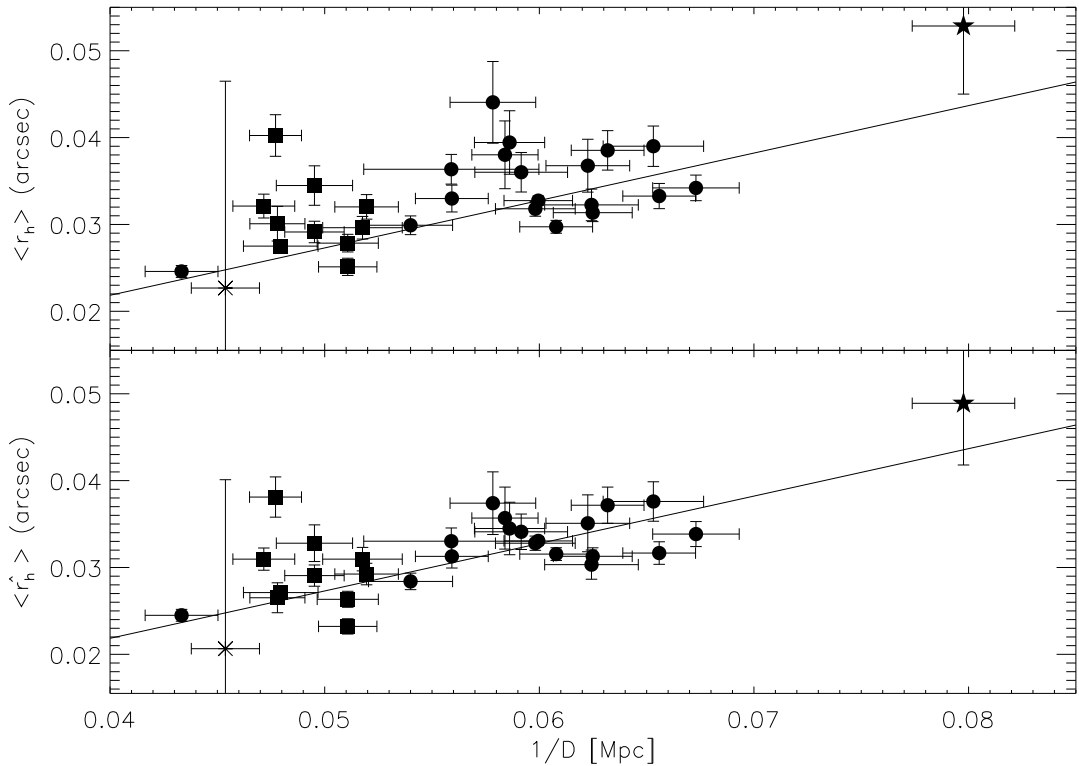}
\caption{Top panel: Average values of $r_h$ (no corrections)
vs. inverse distance for the giant galaxies ($M_B < -19$ mag) in
Fornax (squares), Virgo (circles). Also includes NGC 4697 (star) and
VCC 575 (asterisk). The lower panel shows the same, but for
average $r_h$ values corrected for dependencies on GC and galaxy
color, and local disk surface brightness. The solid line shows the
calibration of $r_h \propto 1/D$ from J05 (normalized to 0\farcs033 at
D=16.5 Mpc, which is also the average value we find for Virgo giant
galaxies).  \label{distancegiants}}
\end{figure*}

We now consider in more detail the use of mean half-light radius as a
distance indicator. We first examine only the bright early-type
galaxies ($M_B < -19$ mag) in which we have shown that the uncorrected
mean half-light radius is very close to constant. Figure
\ref{distancegiants} shows both uncorrected and corrected average
$r_h$ values for bright galaxies ($M_B < -19$ mag) in Fornax and
Virgo, and also NGC 4697 and VCC 575. The values follow very closely
the expected scaling for a constant mean half-light radius. The line
is not a fit to the data, but simply this relation normalized to a value
of $\langle r_h\rangle = 0'\farcs033$ (solid line) at D=16.5 Mpc which is
the average values found for Virgo giants (both in this work and in J05).
 
A simple comparison of the biweight average of all the average values
of uncorrected $r_h$ in GC systems of bright Fornax and Virgo 
early-type galaxies implies a Fornax distance of 18.6$\pm$1.0 Mpc (based on
a Virgo distance of 16.5 Mpc; Mei et~al 2007). Note that the error estimate includes only the
dispersion in the values of $\langle r_h\rangle$ in both Fornax and
Virgo systems; the systematic error on this from the measurement of
the GC sizes is 2.8 Mpc. The significant outlier in the Fornax
galaxies is FCC 21 which as we discussed before has a peculiar
distribution of $r_h$ values\footnote{We note that FCC~21 would be
  readily identified as a peculiar early-type galaxy in any dataset
  where the $r_h$ of GCs can be measured, as the prominent dust lanes
  would be clearly seen. An $r_h$-derived distance on such a galaxy
  could be potentially affected by a large systematic error.}.
Excluding FCC21 from the averaging for Fornax we find
19.1$\pm0.9\pm2.5$ Mpc. The dispersion in the uncorrected value of
$\langle r_h\rangle$ is $0\farcs0041$ (13\%) for all systems in
Fornax cluster bright early types, but FCC 21 accounts for almost half
of this value - by excluding FCC21 it drops to $0\farcs0031$ (11\%).

Applying the corrections discussed in \S6 and \S7 above (note we
apply our corrections to the GC systems in both Fornax and Virgo) we
find a mean angular size of $0\farcs0293$, with dispersion $0\farcs0041$ for all
bright Fornax early types ($0\farcs0285$ and $0\farcs0031$ excluding
FCC 21) while for Virgo we find $0\farcs0331$ and $0\farcs0032$. This
then implies a Fornax distance of $D=18.6\pm0.9\pm2.5$ Mpc (19.1$\pm0.8\pm2.7$ Mpc
excluding FCC 21) if the distance to Virgo is 16.5 Mpc. The dispersion
in the value of the corrected mean half-light radius in Fornax early
types is 14\% (11\% excluding FCC 21) comparable to the dispersion
found by J05 for Virgo cluster systems.

The above discussion relies on an assumed mean distance to Virgo to
test the reliability of the method. We can also use the SBF distances
to the galaxies in both Virgo and Fornax to provide an updated
calibration of the method. Using a biweight mean of all the corrected
mean half-light radii of giant galaxies in both the Fornax and Virgo
clusters (but excluding FCC 21) we find a distance in Mpc of 
\be D =
\frac{0.561 \pm 0.010 \pm 0.060}{\langle\hat{r}_h\rangle} \rm{Mpc}.  
\ee 
where $\langle\hat{r}_h\rangle$ is in arcseconds. This is based on a
constant value of $\langle\hat{r}_h\rangle = 2.71 \pm 0.05 \pm 0.25$ pc for a
GC with color $(g-z) = 1.2$, in a galaxy with color $(g-z)_{\rm gal}
= 1.5$ and at an underlying surface brightness of $\mu_z = 21$ mag
arcsec$^{-2}$. [Note that one prefers to use {\it uncorrected} $r_h$
values in giant galaxies in both Virgo and Fornax, the result is $\langle
r_h\rangle = 2.84 \pm 0.07 \pm 2.9$ pc, so the calibration is then $D = (0.585
\pm 0.010 \pm 0.060)/\langle r_h\rangle $ Mpc.)]

Finally, we can also use the geometric calibration based on the sizes
of GCs in the Milky Way of $\langle\hat{r}_h\rangle \simeq 2.54 \pm
0.1$pc calculated by J05 to provide a distance to the Fornax cluster
independent of other extragalactic distance ladder steps. Using our
mean corrected $\langle\hat{r}_h\rangle$ for Virgo and Fornax
respectively, we find distances of 15.8$\pm0.7\pm1.4$ Mpc and
17.9$\pm1.1\pm1.8$ Mpc (18.4$\pm1.0\pm1.9$ Mpc excluding FCC21). These
distances are reassuringly close to measurements using other techniques
(see, e.g., the discussion in Blakeslee et~al. 2009). Note that the
quoted errors are statistical, plus the systematic error on measuring
the half-light radii on ACS images. Systematic errors introduced by
comparing the size of GCs in early types to those in the Milky Way can
be estimated at about 15-20\%. We estimate that the best
distance to Fornax from this method is $18.4\pm3.7$ Mpc 
(where FCC21 has been excluded).

\begin{figure*}
\epsscale{1}
\plotone{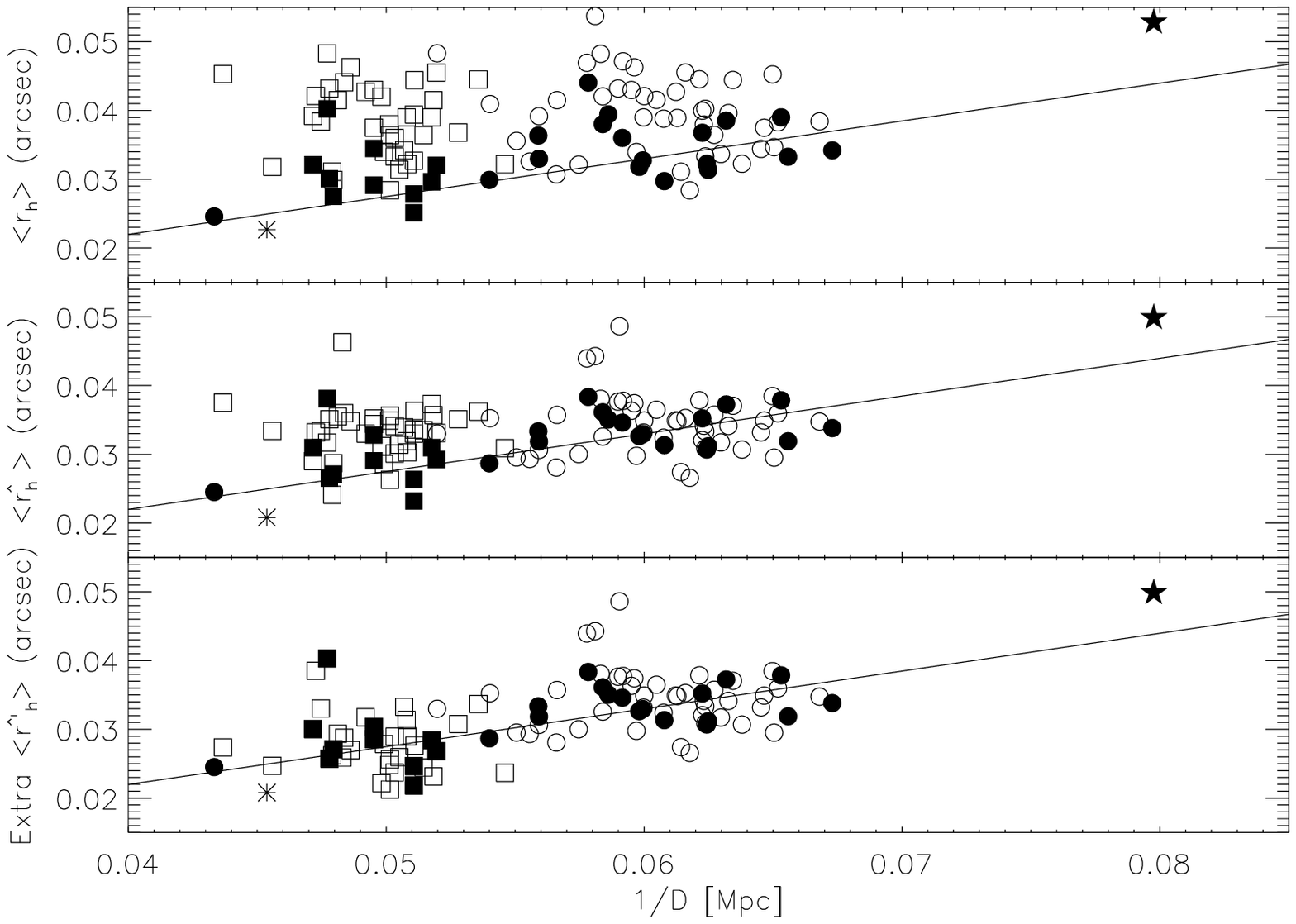}
\caption{As Figure \ref{distancegiants} but now showing both giant and
dwarf galaxies, and also including the extra magnitude correction for
Fornax dwarfs (this is only applied to the GC sizes in Fornax, not in
Virgo, NGC 4697 or VCC 575). Dwarfs are shown as open symbols, giants
as filled. Fornax cluster galaxies are squares, Virgo are circles, NGC
4697 is the star and VCC 575 is the asterisk. We omit error bars
for clarity --- the size of the errors can be seen in
Figure 15.
\label{dwarfs}}
\end{figure*}

We also show in Figure \ref{dwarfs} the same plot for {\it all}
galaxies in Virgo and Fornax together (i.e., also including the fainter
galaxies with smaller numbers of GCs) to illustrate the impact of the
environmental corrections on how the GC systems in the fainter
galaxies fall on the relation. If we use all galaxies together we
estimate a distance to Fornax of $D = 17.2\pm0.6\pm1.9$ Mpc if no
corrections are applied, $D=17.6\pm0.4\pm2.3$ Mpc if the general
corrections (for GC color, local surface brightness at the position
of the GC, and host galaxy color) are applied, and
$D=20.2\pm0.6\pm2.8$ Mpc if the extra correction on host galaxy
luminosity is applied to the sizes of the Fornax cluster galaxy
GCs. We note that this final measurement is very close to the
$20.0\pm1.4$ Mpc recently found by \citet{B09} using the SBF method.
 
Our prescriptions to correct $r_h$ for it to be useful as a standard
ruler depend on the particular observational setup of our program in
that the corrections are given using magnitudes measured in $g$ and
$z$. This limits their usefulness for other observational setups to
some extent, but does not preclude their use. First, as we have shown
above the {\it uncorrected} mean half-light radius is a good distance
indicator for bright galaxies ($M_B < -19$)\footnote{It can be argued
  that a distance needs to be known first in order to to know if
  measuring a distance with uncorrected $r_h$ is appropriate. But
  there other ways the nature of a galaxy as a luminous elliptical
  could be established at the required level of precision: e.g.,~by its
  relative luminosity to other galaxies in the case of a cluster of
  galaxies.}. These are the class of galaxies where obtaining a
distance using $r_h$ would be most useful, as they provide large
number of GCs that are necessary to reduce the statistical uncertainty
in $\langle r_h \rangle$.
In case of a lower luminosity galaxy observed in other bands,
population synthesis models for early-type galaxies and GCs
can be used to transform from our bands to the bandpasses under
consideration. Alternatively, galaxies in well observed galaxy
clusters such as Virgo, or Milky Way GCs, could be used to
infer empirical transformations to the chosen bandpasses if the
appropriate data exist. This is a more involved procedure, but as
pointed out above,  measuring distances using $\langle r_h \rangle$
is not ideally suited for low luminosity galaxies in any
case, due to the small number of GCs generally available in such
galaxies. Finally, we note that our calibration for $\langle r_h \rangle$
hinges on the use of a King model to derive the size of GCs, and
presumably the use of other models could lead to some systematic
differences in the measured half-light radii. Fortunately, it is
possible to calibrate away this potential differences by using the
publicly available data of the ACSVCS galaxies.  All our measured
$r_h$ for GCs belonging to those galaxies have been published in
\citet{J09}. If using a different model translates into a systematic
difference with the $r_h$ measurements presented in \citet{J09}, this
difference can be then be taken into account.

\subsection{Half-Light Radii as Tracers of GC Formation and Evolution}

It is remarkable how little the mean half-light radius of GC systems
varies in early type galaxies in both the Fornax cluster and the Virgo
cluster. The variation is only about 20\% over half a magnitude change
in galaxy color and three orders of magnitude in galaxy luminosity;
when only luminous ($M_B < -19$ mag) galaxies are considered, the variation is
smaller still. This property  clearly provides a
significant constraint on models of GC formation and
evolution. A successful model must also account for the red GCs being $\sim
20$\% smaller than blue ones, for the lack of dependence of GC size on
luminosity (or mass), and for the differing trends of GC half-light
radius with galactocentric distance hinted at in early types (where
the trend is mild) and late types (where it is much steeper).

In J05, two simple models for the formation of GCs were pitted against
some of the observational trends of the half-light radii. The first
was a model in which the half-light radius of the GCs is  determined by overall
pressure confinement of the proto-GC cloud (e.g., McLaughlin \& Pudritz 1996), while a second model
was considered in which the radius is a consequence of tidal limitation of the
GCs. Both models immediately run into a problem of predicting a
scaling of half-light radius with GC mass (or luminosity) which is
significantly larger than the almost zero trend that is observed ($r_h \sim
M^{1/2}$ and $r_h \sim M^{1/3}$, respectively). Thus, in both cases an
unknown mechanism is required --- possibly related to a scaling of star formation
efficiency with cloud binding energy. Assuming such a
mechanism exists, J05 went on to show that in the pressure confined
case there is only a mild dependence of the mean half-light radius
with galaxy luminosity ($\langle r_h \rangle \propto L^{-0.12}$),
while in the tidal limitation model the dependence is exactly zero
(\ie~ $\langle r_h\rangle \sim L_{\rm gal}^0$). They therefore argued
that both models could predict the negligible trend of mean half-light
radius with galaxy luminosity observed in ACSVCS galaxies.

In this paper we observe a slight dependence of $\langle r'_h\rangle$
on galaxy luminosity in both Fornax and Virgo cluster early-types
which clearly differs from the almost zero trend observed in the GC
systems of Virgo cluster early types by J05 (note that slight
differences in the Virgo cluster GC sample, and treatment of local
factors may account for this difference). We observe that for blue GCs
the scaling of size with luminosity is $\langle r'_h \rangle \propto
L^{-0.05\pm0.02}$ and for red GCs it is $\langle r'_h \rangle \propto
L^{-0.11\pm0.04}$. A possible explanation of the difference between
the subpopulations could be that as the red GCs tend to be
concentrated more towards the centre of the galaxy they may be more
significantly impacted by the size of the galaxy (\ie ~the depth of
its potential well). Overall, this evidence of a mild galaxy luminosity
dependence of GC sizes, slightly favours the pressure confinement
scenario over the tidal limitation model, especially for the red
subpopulation, however since the trends are so marginal, and appear to
depend on the GC selection procedure, we do not consider this strong
evidence.

We also see some small evidence for a shallower trend of the GC sizes
with local surface density of the host galaxy in the Fornax galaxies
than was observed in the Virgo galaxies as well as a slightly
shallower trend with GC color (once global and local correlations are
corrected for). Such subtle differences in the environmental
dependencies of the sizes of the GCs in Fornax cluster galaxies versus
Virgo cluster galaxies may be hinting at a dependence of the formation
mechanisms or more likely the structural and dynamical evolution of
GCs on the global environment of the host galaxies.

As in J05 the slope of the relation between half-light radius and
galactocentric distance ($b_r$) is significantly lower than is found
for GCs in the Milky Way (MW), which (after projection effects are
considered) follow a relation with $b_{r,{\rm MW}} \sim 0.3 \pm
0.15$. This study then further confirms the discrepancy between the
galactocentric trends in early types and in the MW, however we note
that our selection removes GCs with $r_h>10$pc (as discussed in
\citealt{J09}) which could be driving the large trend seen in the MW
GC system. If real, this difference in the GC systems is clearly of
interest for further study and may point to a difference in the
formation or dynamical evolutions of GCs in early type galaxies
vs. late type galaxies or, possibly, of galaxies in low density
environments vs. clusters.

There have been other recent papers which take advantage of the ACS on
HST to measure the half-light radii of GCs in galaxies outside the LG
and study their behaviour with galactocentric distance. \citet{DG07}
show that the GCs in NGC 1533 (an SB0 galaxy in the Dorado group which
appears to be transitioning from late to early type) have a $r_h$
vs. galactocentric radius relation more similar to the GCs in the
Milky Way than in Virgo or Fornax cluster early type
galaxies. \citet{S06} study the GC system of the Sombrero galaxy, an
Sa or S0 galaxy also in the transition region between late and early
types and see similar behavior (they find $b_{\rm blue} = 0.16\pm0.04$,
$b_{\rm red} = 0.32\pm0.05$ and $b_{\rm all} = 0.24\pm0.03$ using the
same notation we use above). \citet{H09a} studied GC candidates in NGC
891 (a nearby edge-on spiral thought to be similar to the MW) and
while the numbers of GCs in NGC 891 are quite small the noisy trend
seen with galactocentric radius seems consistent with that seen in the
MW. These studies are indicative of a general property of the GC
systems in late-type galaxies in small groups to have a steeper
dependence on galactocentric radius than is seen in the systems of
early-type galaxies in clusters however the statistics here are still
very small.  It would be interesting to study the GC systems of more
late-type galaxies to test this hypothesis. At the moment, it cannot be
determined if it is the galaxy morphology or, the environment in which the galaxy resides,
that drives this change. A study of GC systems of early types in
low density environments and late types in higher density environments
would be instructive to disentangle the effects of morphology and
density. \citet{DG07} tentatively explain the difference as the effect
of steeper density gradients in small groups like Dorado or the LG
versus large clusters. Since the sizes of GCs may be limited by tidal
fields, the argument is that GC sizes should have a stronger dependence
on radius in small groups than in clusters. We find here a
(marginally) smaller trend of GC radius on galactocentric radius for
the GC systems in Fornax cluster galaxies than in the Virgo cluster
and while Virgo is the larger cluster, the central density of Fornax
is about twice that of Virgo in term of galaxies per Mpc$^3$ (see
Table 1 of \citealt{J07a}). Based
on this measure of density we may then be observing the trend expected by the \citet{DG07} model.
 
It is also interesting that while the size-galactocentric radius
relation appears to point to differences in the GC systems of
different Hubble types (or galaxies in difference environments),
recent studies of the dispersion of the GC luminosity function (GCLF)
instead favour the idea that GC systems in late type galaxies
(specifically the Milky Way and M31) are not different from those in
early types since they fit on the same dispersion-luminosity relation
of systems in early types \citep{J07b}. The color distribution of the
GC systems of late and early types have also been shown to be similar
and have similar trends with galaxy luminosity \citep{P06}. At the
moment the only evidence suggesting a significant difference between
the GC systems in late and early type galaxies is the change in the
size-galactocentric radius relation. Clearly further work, including
studies to significantly greater galactocentric radii need to be done
to confirm this as a general property.

\section{Conclusions}

We have used data on the half-light radii of GCs belonging to galaxies
observed in the ACSFCS to extend the work of J05 in which the
environmental dependencies of the half-light radii of GCs in early-type
galaxies in the ACSVCS were studied, and a corrected mean
half-light radius was suggested as a reliable distance indicator.  By
adding data from the ACSFCS, we increase the sample size for the study
of the environmental dependencies, and add leverage to the study of
the corrected half-light radius as a possible distance indicator
(since Fornax is at a larger distance than the Virgo cluster).

We find only subtle differences in the environmental dependencies of
the sizes of GCs in Fornax cluster galaxies from what was found in the
Virgo cluster by J05. In addition, we perform a
Principal Component Analysis (PCA) to check that no major correlations
are being hidden (in Appendix A).

Looking at 2D relations, we again confirm the well known results that
there is no correlation between GC size, $r_h$, and mass (for $M
< 2\times 10^6 M_\odot$, but that blue GCs ($(g-z) < 1.05$ with
$\langle r_h\rangle = 3.36\pm 0.03\pm0.25$ pc) are about 20\% larger than
red ones ($(g-z) > 1.15$ with $\langle r_h\rangle = 2.83 \pm
0.02\pm0.25$ pc).

We show that the half-light radii of GCs in early-type galaxies in the
Fornax cluster increase only slightly with galactocentric radius (or
decreasing surface brightness) as was also found by J05 for systems in
Virgo cluster early-types. In fact, the trend we find in Fornax cluster
systems is slightly shallower than that seen in the Virgo cluster.  As
was found in J05, the trend of $r_h$ with galactocentric distance is
significantly shallower (only $2\sigma$ different from zero in Fornax
galaxy GC systems) than that observed in our Galaxy perhaps pointing
to a different formation scenario, or evolution of GCs in early types
vs. late types (or in high density regions vs. low density
regions). We discuss this observation in light of other recent studies
of the GC systems of late-type galaxies and argue there is now some
evidence for a general property of the GC systems of late-type
galaxies in small groups to have a stronger trend of half-light radius
with galactocentric distance than is seen in early types in
clusters. This will provide stronger constraints on the differences in
formation and evolution of GCs in different types of galaxies in
different environments.

We confirm the trend of mean half-light radius with galaxy color that
was first observed in J05, but show suggestions that there is a
residual correlation with galaxy luminosity in the mean half-light
radius of GC systems of Fornax early-type galaxies which is larger
than that seen in GC systems of Virgo early types.  We revisit the two
simple pictures of the origin of $r_h$ in GCs discussed by J05,
arguing that the additional trend we observe for the mean half-light
radius of GC systems in both Fornax and Virgo cluster early type
galaxies to decrease with galaxy luminosity (as $\langle r'_h \rangle
\propto L^{-0.05\pm0.02}$ in blue GCs and $\langle r'_h \rangle
\propto L^{-0.11\pm0.04}$ in red GCs) may be providing some support
for the ``pressure-confined proto-GC cloud'' model over tidal
limitation of $r_h$. However since the size of this trend appears to
depend on the details of the GC selection (as it was not present in
the similar J05 sample) we suggest that it should not be over
interpreted.

We show that for the most luminous galaxies ($M_B < -19$ mag), the uncorrected
mean half-light radius is by itself an excellent distance indicator,
varying by around 10-15\% across galaxies. This is especially true if
we remove the unusual GC system of FCC 21 (Fornax~A). Once corrected
for dependencies on GC and galaxy color and local surface brightness
we find a constant value of $\langle\hat{r}_h\rangle = 2.71 \pm
0.05\pm0.25$ pc for a GC with color $(g-z) = 1.2$, in a galaxy with
color $(g-z)_{\rm gal} = 1.5$ and at an underlying surface brightness
of $\mu_z = 21$ mag arcsec$^{-2}$ across giant galaxies in both the
Virgo and Fornax clusters (for Virgo alone we find
$\langle\hat{r}_h\rangle = 2.67 \pm 0.07\pm0.25$ pc, in Fornax it is
$\langle\hat{r}_h\rangle = 2.78 \pm 0.12\pm0.25$ pc excluding
FCC21). The same simple geometric calibration used by J05 to estimate
an independent distance to the Virgo cluster of $D_{\rm Virgo} =
16\pm2.3$ Mpc gives a distance to the Fornax cluster of $D_{\rm
  Fornax} = 18.4\pm3.7$ Mpc (excluding GCs in FCC 21).

This extension of the work of J05 to include GC systems in early-type
galaxies in the Fornax cluster adds support to the idea of a constant
mean $r_h$ in luminous early type galaxies, but suggests that the
environmental dependencies may be subtly different in different
environments which is especially important in the lower mass
galaxies. The mean half-light radii in GC systems of massive early-type 
galaxies has the potential to provide a geometric distance
measurement to bright early-type galaxies, which could reach cosmologically
interesting distances (as quantified by J05) in the era of giant optical
telescopes with adaptive optics. 

\begin{acknowledgements}
Support for programs GO-10217 and GO-9401 was provided through grants
from the Space Telescope Science Institute, which is operated by the
Association of Universities for Research in Astronomy, Inc., under
NASA contract NAS5-26555. K.~L.~M. acknowledges funding from the Peter
and Patricia Gruber Foundation as the 2008 Peter and Patricia Gruber
Foundation International Astronomical Union Fellow, and from the
University of Portsmouth. A.~J. and L.~I. acknowledge support from the
Chilean Center of Excellence in Astrophysics and Associated
Technologies PFB-06, and from the Chilean Center for Astrophysics
FONDAP 15010003. A.~J. acknowledges additional support from MIDEPLAN
ICM Nucleus P07-021F.
   
\end{acknowledgements}

\appendix
\section{Principal Component Analysis}

\begin{deluxetable}{lccccccccc|c}
\tablewidth{0pc} 
\tablecaption{PCA results for input of all 8 environmental factors}
\tablehead{\colhead{Factor} & \colhead{V1} &  \colhead{V2} &  \colhead{V3} &  \colhead{V4} &  \colhead{V5} &  \colhead{V6} &  \colhead{V7}&  \colhead{V8}&  \colhead{V9}& \colhead{$\sigma$} \label{pca9}}
\startdata 
$\log r_h$          &  0.15$\pm$0.02 &  0.18$\pm$0.02 &  0.69$\pm$0.03 &  0.17$\pm$0.08 &  0.40$\pm$0.11 &  0.47$\pm$0.10 &  0.13$\pm$0.05 &  0.07$\pm$0.02 &  0.02$\pm$0.01& 0.157\\
$(g-z)_{\rm GC}$    & -0.29$\pm$0.02 & -0.27$\pm$0.02 & -0.29$\pm$0.05 & -0.25$\pm$0.11 & -0.11$\pm$0.18 &  0.72$\pm$0.11 &  0.29$\pm$0.07 & -0.16$\pm$0.02 &  0.03$\pm$0.01& 0.231\\
$z_{\rm GC}$        &  0.13$\pm$0.02 &  0.12$\pm$0.02 & -0.49$\pm$0.10 &  0.68$\pm$0.33 &  0.18$\pm$0.09 &  0.26$\pm$0.07 & -0.17$\pm$0.05 & -0.01$\pm$0.01 & -0.01$\pm$0.01& 0.794\\
$\log (r_p/r_e)$    &  0.56$\pm$0.01 & -0.21$\pm$0.03 &  0.05$\pm$0.02 & -0.07$\pm$0.04 & -0.23$\pm$0.05 &  0.14$\pm$0.07 & -0.17$\pm$0.03 & -0.15$\pm$0.01 & -0.71$\pm$0.01& 0.384\\
$\mu_z$ at $r_{p}$  &  0.53$\pm$0.01 &  0.05$\pm$0.03 & -0.10$\pm$0.04 & -0.28$\pm$0.12 & -0.08$\pm$0.07 &  0.21$\pm$0.05 & -0.42$\pm$0.07 &  0.10$\pm$0.02 &  0.59$\pm$0.01& 1.657\\
$\mu_g-\mu_z     $  &  0.25$\pm$0.02 & -0.25$\pm$0.02 & -0.29$\pm$0.06 & -0.26$\pm$0.12 &  0.75$\pm$0.13 & -0.16$\pm$0.19 &  0.20$\pm$0.06 & -0.13$\pm$0.01 & -0.04$\pm$0.01& 1.413\\
$M_{B,{\rm gal}}$   &  0.40$\pm$0.01 &  0.29$\pm$0.02 & -0.17$\pm$0.02 &  0.00$\pm$0.04 & -0.22$\pm$0.05 & -0.04$\pm$0.08 &  0.71$\pm$0.11 &  0.39$\pm$0.02 &  0.03$\pm$0.01& 1.373\\
$(g-z)_{\rm gal}$   & -0.06$\pm$0.03 & -0.64$\pm$0.01 &  0.06$\pm$0.02 &  0.12$\pm$0.05 &  0.03$\pm$0.02 &  0.01$\pm$0.02 & -0.08$\pm$0.02 &  0.74$\pm$0.01 & -0.00$\pm$0.01& 0.092\\
$\langle \mu_z \rangle$           & -0.22$\pm$0.03 &  0.52$\pm$0.02 & -0.22$\pm$0.05 & -0.32$\pm$0.13 &  0.20$\pm$0.05 &  0.15$\pm$0.06 & -0.29$\pm$0.05 &  0.47$\pm$0.01 & -0.38$\pm$0.01& 0.927\\
\hline
E-values (\%) &  28.0$\pm$  0.3 &  21.6$\pm$  0.4 &  12.0$\pm$  0.2 &  10.7$\pm$  0.2 &   9.0$\pm$  0.2 &   8.2$\pm$  0.2 &   6.5$\pm$  0.2 &   3.3$\pm$  0.1 &   0.7$\pm$  0.1\\
$E_{i-1}/E_i$ &  &   1.3 &   1.8 &   1.1 &   1.2 &   1.1 &   1.3 &   2.0 &   4.8
\enddata
\end{deluxetable}

Principal component analysis (PCA) provides a statistical method to
search for correlations between two or more correlated
variables. Since (as described in \S4) the size of a GC may
depend on several inter-related properties of the host galaxy, local
galaxy environment and of the GC itself the problem is an ideal
candidate for a PCA. PCA decomposes the observed correlations between
parameters into a set of eigenvectors and eigenvalues describing the
main variance seen in the data. This method is now becoming a standard
in astronomical applications where intercorrelations between several
variables are common, so we do not provide a full explanation of the
method here, but rather refer the reader to standard references below.
Here we use the PCA routine in the Astronomy IDL Library\footnote{\tt http://idlastro.gsfc.nasa.gov/} (which
performs PCA according to the method described in ``Multivariate Data
Analysis" by \citealt{MH87}). We follow closely the method described
in \citet{W08} in the interpretation of the PCA results and in the use
of bootstrap re-sampling to calculate the errors.

 The goals of our PCA will not be to derive expressions for the
dependence on $r_h$ on other factors, but rather to look at the shape
of the trends we expect to see and which relations we expect to be
important and unimportant. In a sense, we use this to arrive at the
optimal number of parameters to describe the $r_h$
variations. Traditional 2D relations are easier to interpret so we will
return to them in the main body of the paper (\S5) once we understand the overall shape of the expected
interdependencies.

We initially perform a PCA on Fornax cluster GCs using all 8
environmental factors described in Section 4 (in log, or
magnitude space) as well as $\log r_h$. Numerical results are shown in
Table \ref{pca9}. In this table (and Tables A2, A3 and A4 below) the
eigenvector outputs (describing the directions of the main variance in the data) are labelled at V$i$ (for a PCA with $n$
variables, $i=1..n$). We list the projection of each input variable
onto these eigenvectors as well as the eigenvalue (``E-value")
expressed as a percentage of the total.  We also show the ratios
between successive eigenvalues, $E_{i-1}/E_i$. The PCA routine
performs analysis on standardized variables, meaning that they are
rescaled to have a zero mean and unit variance. This means that the
relative sizes of the eigenvalues tells something physical about the
relative variance of the data in different ``directions''. The last
column, labelled $\sigma$, shows the total variance in the input
variable before it was rescaled.

We can see in Table A1 that the first two eigenvalues are comparable in
size ($E_1/E_2 = 1.3$) and about twice as large as the third, after
which there are 5 vectors of comparable size (with ratios of $\sim
1.0-1.3$), until the last two values which are smaller again. In
physical terms this can be interpreted as the variance in the data
being slightly larger in 2 directions, then of comparable size until
we have two directions which have quite small variance. There there is
not one {\it much} stronger trend between variables than any
other. From the size of the components on the first eigenvector we see
that the strongest trend is between the projected radius ($\log
(r_p/r_e)$), the local surface brightness, $\mu_z$ (since these have
the largest components of the first eigenvector). The next strongest
trend (two largest projections onto V2) is between the galaxy color
and average surface brightness.

Obviously $\mu_z$ and $r_p/r_e$ are related through the surface
brightness profile of the galaxy, and therefore a relatively strong
trend between them is expected. Since these trace almost the same
physical property (i.e., the local environment of the GCs) we chose to
drop $r_p/r_e$ in further analysis (this choice is also discussed
in Section 5).

\begin{deluxetable}{lcccccc|c}
\tablewidth{0pc} 
\tablecaption{PCA results for input of reduced set of 5 environmental factors}
\tablehead{\colhead{Factor} & \colhead{V1} &  \colhead{V2} &  \colhead{V3} &  \colhead{V4} &  \colhead{V5}&  \colhead{V6}  & \colhead{$\sigma$} \label{pca6}}
\startdata 
$\log r_h$           &  0.27$\pm$0.03 &  0.29$\pm$0.04 &  0.62$\pm$0.04 &  0.66$\pm$0.03 &  0.08$\pm$0.05 &  0.06$\pm$0.02& 0.157\\
$(g-z)_{\rm GC}$     & -0.34$\pm$0.03 & -0.45$\pm$0.03 & -0.26$\pm$0.03 &  0.64$\pm$0.03 & -0.24$\pm$0.23 & -0.28$\pm$0.02& 0.231\\
$\mu_z$ at $r_{p}$   &  0.16$\pm$0.05 &  0.59$\pm$0.02 & -0.35$\pm$0.03 &  0.06$\pm$0.06 & -0.54$\pm$0.43 &  0.12$\pm$0.02& 1.657\\
$\mu_g-\mu_z$        & -0.25$\pm$0.05 &  0.49$\pm$0.03 & -0.48$\pm$0.03 &  0.28$\pm$0.06 &  0.48$\pm$0.37 & -0.12$\pm$0.02& 1.413\\
$(g-z)_{\rm gal}$    & -0.67$\pm$0.01 &  0.05$\pm$0.05 &  0.17$\pm$0.02 &  0.03$\pm$0.02 & -0.05$\pm$0.05 &  0.72$\pm$0.01& 0.092\\
$\langle \mu_z \rangle$            &  0.52$\pm$0.02 & -0.33$\pm$0.04 & -0.40$\pm$0.02 &  0.24$\pm$0.03 &  0.14$\pm$0.10 &  0.61$\pm$0.01& 1.199\\
\hline
E-values (\%) &  29.3$\pm$0.5 &  24.8$\pm$0.4 &  17.0$\pm$0.4 &  12.8$\pm$0.3 &  10.3$\pm$0.3 &   5.8$\pm$0.2\\
$E_{i-1}/E_i$ &  &  1.2 &  1.5 &  1.3 &  1.2 &  1.8
\enddata
\end{deluxetable}

Running PCA on the remaining 7 environmental factors and the
half-light radius shows that the strongest residual correlations are
between $M_{B,{\rm gal}}$ and the color of the galaxy. We argue below
that $M_{B,{\rm gal}}$ and $(g-z)_{\rm gal}$ are providing
complementary information about the host galaxy (because of the well
known color-magnitude relation in early-type galaxies), and therefore
chose to keep galaxy color as a distance independent variable. We
will also drop the GC magnitude $z$ at this point as the last
remaining distance dependent factor.

Running the PCA on the remaining 5 environmental factors we then find
a distribution which shows a relatively strong trend between the
galaxy color and surface brightness (both local and global), and
weaker trends between the rest of the variables. The half-light radius
($\log(r_h)$) is most strongly projected onto the third eigenvector
along with (in decreasing order) the color at the position of the GC
($\mu_g - \mu_z$) and the surface brightness at position of the GC
($\mu_z$) --- both tracers of the local environment of the GC. The full
numeric outcome is shown in Table \ref{pca6}.

We run two final PCA on a set of 3 environmental factors plus the GC
half-light radius - one in which the global and local environment
properties are described by surface brightness, and another using the
global and local colors. These final PCA on a limited set of
parameters now indicate a roughly ``spherical'' distribution of
points, in both of these 4-D spaces (see numerical results in
Tables \ref{pca2} and \ref{pca22}). This indicates that there is no one trend between
parameters which is significantly stronger than any other. It also seems to
not matter if we use the surface brightness or color to trace of the global and local properties of the host galaxy.

\begin{deluxetable}{lcccc|c}
\tablewidth{0pc} 
\tablecaption{PCA results for input of 1st reduced set of 3 environmental factors - using colors}
\tablehead{\colhead{Factor} & \colhead{V1} &  \colhead{V2} &  \colhead{V3} &  \colhead{V4} & \colhead{$\sigma$} \label{pca2}}
\startdata 
$\log r_h$                  &  0.49$\pm$0.02 &  0.28$\pm$0.06 &  0.81$\pm$0.02 &  0.14$\pm$0.05 & 0.158\\
$(g-z)_{\rm GC}$            & -0.56$\pm$0.02 & -0.40$\pm$0.04 &  0.37$\pm$0.04 &  0.61$\pm$0.09 & 0.231\\
$\mu_g-\mu_z$ at $r_{p}$    & -0.27$\pm$0.04 &  0.84$\pm$0.02 & -0.20$\pm$0.06 &  0.42$\pm$0.07 & 1.657\\
$(g-z)_{\rm gal}$           & -0.61$\pm$0.01 &  0.22$\pm$0.03 &  0.39$\pm$0.05 & -0.64$\pm$0.10 & 0.092\\
\hline
E-values (\%) &   37.0$\pm$0.6 &   26.3$\pm$0.4 &   20.8$\pm$0.4 &   15.9$\pm$0.4\\
$E_{i-1}/E_i$ &  &    1.4 &    1.3 &    1.3
\enddata
\end{deluxetable}

\begin{deluxetable}{lcccc|c}
\tablewidth{0pc} 
\tablecaption{PCA results for input of 2nd reduced set of 3 environmental factors - using surface brightness}
\tablehead{\colhead{Factor} & \colhead{V1} &  \colhead{V2} &  \colhead{V3} &  \colhead{V4} & \colhead{$\sigma$} \label{pca22}}
\startdata 
$\log r_h$            & -0.50$\pm$0.13 & -0.20$\pm$0.10 & -0.52$\pm$0.52 & -0.25$\pm$0.30 & 0.158\\
$(g-z)_{\rm GC}$      &  0.62$\pm$0.16 &  0.06$\pm$0.05 & -0.05$\pm$0.07 & -0.50$\pm$0.58 & 0.231\\
$\mu_z$ at $r_{p}$    & -0.54$\pm$0.14 &  0.03$\pm$0.09 &  0.46$\pm$0.45 & -0.34$\pm$0.39 & 1.657\\
$\langle \mu_z \rangle$              &  0.13$\pm$0.05 & -0.96$\pm$0.12 &  0.16$\pm$0.13 &  0.01$\pm$0.06 & 0.927\\
\hline
E-values (\%) &   35.0$\pm$  0.5 &   25.0$\pm$  0.2 &   22.0$\pm$  0.4 &   18.1$\pm$  0.4\\
$E_{i-1}/E_i$ &  &    1.4 &    1.1 &    1.2
\enddata
\end{deluxetable}

 The scaling relation between variables $X$ and $Y$ can be found from
 the primary eigenvector, rescaled by the standard deviations of $X$
 and $Y$ as \be X \propto \frac{V_1(Y)/\sigma_Y}{V_1(X)/\sigma_X} Y.
 \ee We do not expect to find exactly the same scaling relations of
 $\log r_h$ here as we will when we consider the 2D trends since
 in the 2D trends we attempt to find a representative fit not simply
 dominated by the massive galaxies with large numbers of GCs (and
 generally at smaller radii compared to $r_e$ because of the size of
 the ACS field). For example below we often fit the trends in each
 galaxy separately, then make averages of all the trends in all the
 galaxies. We also consider blue and red GCs separately (although find
 no significant differences), and consider the trend of $\log r_h$
 with GC color only after the corrections for local and global
 factors are made. All of this is explained in greater detail in Section 5
  but is included here to explain why we do not place
 weight on the scaling relations found from the PCA, but rather use it
 only as a way to make sure we are not missing any unexpected
 correlations and have included all the important factors.

 The final conclusion from our PCA here is that the variability in the
 values of $r_h$ of GCs can be described reasonably well using 3
 factors, one each from the internal, local and global properties (as
 defined in \S4).

\end{document}